\documentclass[%
 reprint,
superscriptaddress,
 amsmath,amssymb,
 aps,
]{revtex4-2}

\usepackage{graphicx}
\usepackage{dcolumn}
\usepackage{bm}
\usepackage[utf8]{inputenc}
\usepackage{float}
\usepackage{textcomp}
\usepackage{amsmath}
\usepackage{url}
\usepackage{grffile}
\usepackage{bbold}
\usepackage{color}
\usepackage{gensymb}
\DeclareGraphicsExtensions{.png .jpg .pdf}
\usepackage{hyperref}
\hypersetup{
     colorlinks = true,
     linkcolor = blue,
     anchorcolor = blue,
     citecolor = blue,
     filecolor = blue,
     urlcolor = blue
     }
\usepackage{braket}
\usepackage{physics}
\usepackage{array}
\usepackage{booktabs}
\usepackage{soul}
\usepackage{chemformula}
\usepackage[version=4]{mhchem} 
\usepackage{comment}
\usepackage{natbib}
\usepackage{caption}
\usepackage{subcaption}

\usepackage{pdfpages} 
\usepackage{pgffor} 
\usepackage{xr} 

\makeatletter
\AtBeginDocument{\let\LS@rot\@undefined}
\makeatother

\def\supplementfilename{Supp_GKA_arxiv}

\pdfximage{\supplementfilename.pdf}
\def\numbersupplementpages{\the\pdflastximagepages}

\newif\ifarXiv
\arXivtrue 

\footnotetext{$^\dagger$ These authors contributed equally to this work.}
\footnotetext{$^*$ Email: milorad.milosevic@uantwerpen.be}

\begin{document}

\preprint{APS/123-QED}

\title{Goodenough-Kanamori-Anderson high-temperature ferromagnetism in tetragonal transition-metal xenes}

\author{U. Yorulmaz$^\dagger$}
\affiliation{Department of Physics, Eskisehir Osmangazi University, Eskisehir, Turkiye}
\affiliation{Department of Physics \& NANOlab Center of Excellence, University of Antwerp, Groenenborgerlaan 171, B-2020 Antwerp, Belgium}

\author{D. \v{S}abani$^\dagger$}
\affiliation{Department of Physics \& NANOlab Center of Excellence, University of Antwerp, Groenenborgerlaan 171, B-2020 Antwerp, Belgium}

\author{C. Sevik}
\affiliation{Department of Mechanical Engineering, Eskisehir Technical University, Eskisehir, Turkiye}
\affiliation{Department of Physics \& NANOlab Center of Excellence, University of Antwerp, Groenenborgerlaan 171, B-2020 Antwerp, Belgium}

\author{M. V. Milo\v{s}evi\'c$^{*}$}
\affiliation{Department of Physics \& NANOlab Center of Excellence, University of Antwerp, Groenenborgerlaan 171, B-2020 Antwerp, Belgium}

\date{\today}

\begin{abstract}
Seminal Goodenough-Kanamori-Anderson (GKA) rules provide the inceptive understanding of the superexchange interaction of two magnetic metal ions bridged with an anion, and suggest fostered ferromagnetic interaction for orthogonal bridging bonds. However, there are no examples of two-dimensional (2D) materials with structure that optimizes the GKA arguments towards enhanced ferromagnetism and its critical temperature. Here we reveal that an ideally planar GKA ferromagnetism is indeed stable in selected tetragonal transition-metal xenes (tTMXs), with Curie temperature above 300~K found in CrC and MnC. We provide the general orbitally-resolved analysis of magnetic interactions that supports the claims and sheds light at the mechanisms dominating the magnetic exchange process in these structures. With recent advent of epitaxially-grown tetragonal 2D materials, our findings earmark tTMXs for facilitated spintronic and magnonic applications, or as a desirable magnetic constituent of functional 2D heterostructures.
\end{abstract}

\maketitle
\section{Introduction}
The experimental discovery of the premiere magnetic two-dimensional materials (M2DMs) - CrI$_{3}$ \cite{CrI32017} and CrGeTe$_{3}$ \cite{CrGeTe2017,CrGeTe2017_Xing} - opened the floodgates to many emergent 2D materials of this class. Numerous theoretical \cite{Lado2017,CrX32018,Kitaev2018,Kashin2020} and experimental studies \cite{FGT2018_E,CrI3_dop,CrI3_E,VSe2} followed, explaining the origins and possible manipulations of the long-range magnetic order in the monolayer limit. It is needless to emphasize that intrinsically room-temperature M2DMs would be highly beneficial for applications in sensing, spintronics, and otherwise, and bear promise towards high tunability by diverse mechanical, chemical, and electronic means. However, it quickly became clear that critical temperatures (T$_\textrm{c}$) for magnetic order to vanish are by rule always smaller in 2D materials compared to their bulk counterparts \cite{CrI32017,FGT2020,NiI2_multiferoic}. Namely, in order to host sizable regions with magnetic order (and circumvent limitations imposed by the Mermin-Wagner theorem \cite{MerminWagner}), M2DMs require anisotropy in magnetic exchange. That needed anisotropy is known to originate from the spin-orbit coupling (SOC), which is in general much weaker compared to e.g. Coulomb attraction or repulsion of charged particles, hence attains very low magnitudes (of typically 0.01-0.1~meV). Another reason for T$_\textrm{c}$ to decrease as one thins the magnetic material from bulk to a monolayer stems from the correspondingly diminishing magnetic exchange along the third, out-of-plane direction. One may therefore expect that magnetic order in 2D materials is strictly limited to the very low T$_\textrm{c}$, but that is not necessarily the case. For example, Fe$_{3}$GeTe$_{2}$ hosts the ferromagnetic (FM) order up to 130~K in the monolayer (ML) limit \cite{FGT2018}. In addition, the FM order and T$_\textrm{c}$ of 213~K were measured in few-layer thick 1T-CrTe$_{2}$ \cite{1TCrTe22021}, with the unusual trend that T$_\textrm{c}$ increases as one goes from bulk to few-layers. Furthermore, in a few-layer FePS$_{3}$, the antiferromagnetic (AFM) order was observed up to the T$_\textrm{c}$ of 120~K \cite{FePS2021}.
Finally, FM order was measured even at room temperature in thicker films~($\approx10$ nm, see e.g. \cite{fewL1,fewL2}), but also in monolayer MnSe$_x$~\cite{MnSex,FMinMnSSe2}, VSe$_2$~\cite{VSe2,originVSe2}, and Cr$_3$Te$_4$~\cite{Cr3Te4}. The latter samples were deposited epitaxially, which in general involves structural defects \cite{originVSe2} and strong interfacial effects with the substrate \cite{originVSe2,Cr3Te4} into the origins of the observed robust magnetic interactions, which complicates theoretical interpretations. Otherwise, the isotropic magnetic interactions in crystalline monolayers are relatively straightforward to extract theoretically in all available first-principles codes. Such studies, on predominantly \textit{in silico} created 2D materials, have yielded many predictions of high- or even room-temperature intrinsic magnetism~\cite{roomT1,roomT2,roomT3,roomT4,roomT5,FMinMnSSe2}. However, those predictions typically failed to quantify the underlying microscopic mechanisms for such a large predicted magnetic exchange. 
\begin{figure}[t]
\centering
\includegraphics[width=\linewidth]{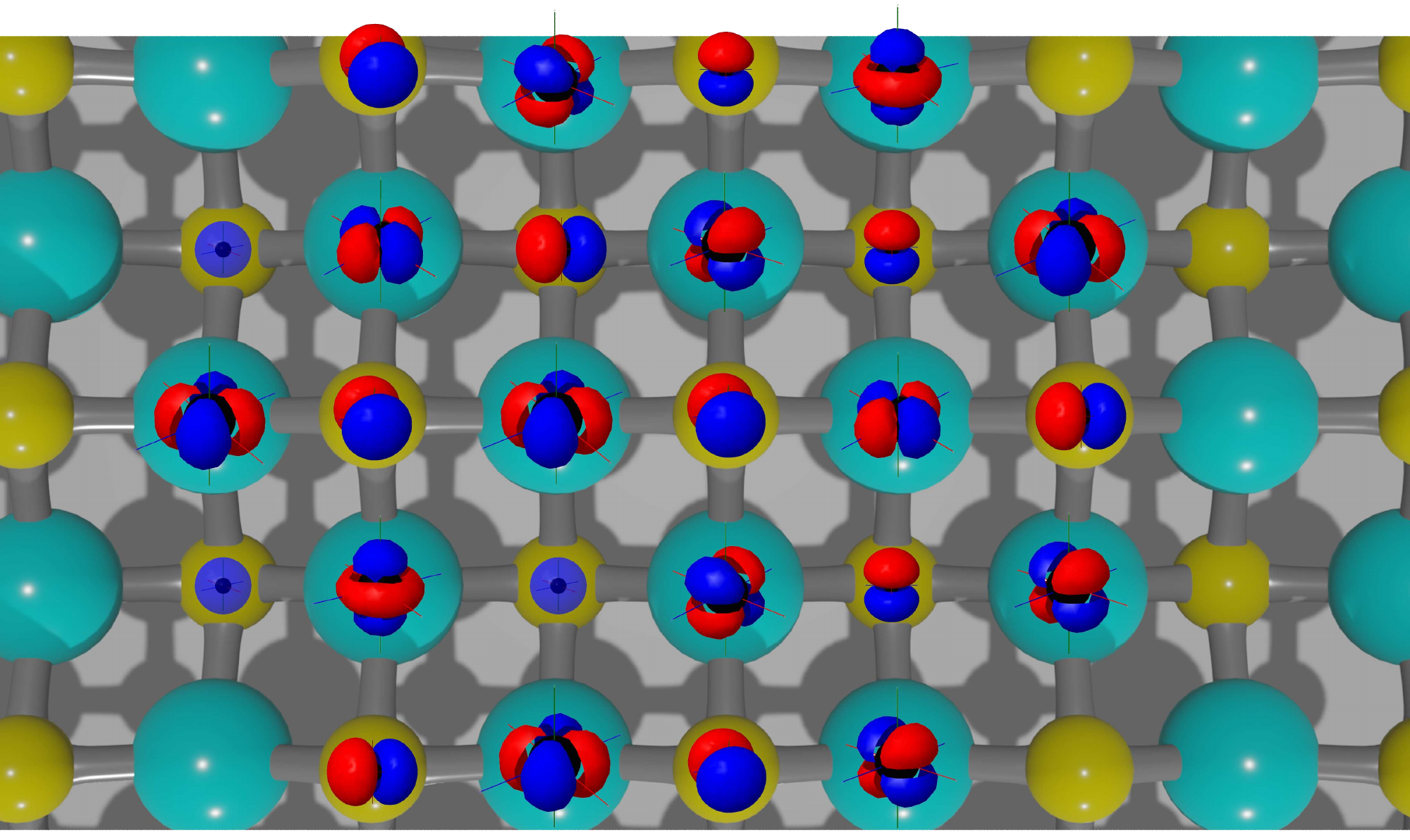}
\caption{\textbf{Crystal structure of a monolayer tTMX.} TM (cyan) atoms are sketched with one $d$ orbital, and X (yellow) atoms with one $p$ orbital, since these orbitals and their hybridization are essential for magnetic interactions in this system.}
\label{figure1}
\end{figure}

In this paper we therefore take a step back, and explore the route towards room-temperature 2D ferromagnetism starting from the well-established set of Goodenough-Kanamori-Anderson (GKA) empirical theoretical rules from the late 1950s~\cite{goodenough,kanamori,anderson1,anderson2}. We accordingly aim at monolayers with 90$^\circ$ between the nearest magnetic atoms (A) connected by a ligand (X), thus ideally a \textit{Lieb-lattice material} of A$_2$X type. However, a magnetic 2D material of such specific planar structure has not been reported to date, although some Lieb-lattice 2D materials have been considered computationally for other purposes (see e.g. Ref.~\cite{nanoresearch14}). As a best available choice, for not only geometry but also sizable SOC, we instead focus on the family of tetragonal transition-metal xenes (tTMXs; see Fig.~\ref{figure1}), seeking a square-lattice planar material among them - still with 90$^\circ$ TM-X-TM nearest-neighbor bonds. 
Using advanced methodology on top of the standard first-principles approaches based on Density Functional Theory (DFT), we computationally validate the structural stability and strong intrinsic magnetic interactions in these materials, detail the microscopic (orbital-resolved) origin of enhanced magnetic exchange, and identify CrC and MnC as premiere square-lattice monolayer ferromagnets with a particularly high T$_\textrm{c}$.

\section{Results}
We commence our analysis with a computationally crude throughput screening of dynamical stability and magnetic interactions in tTMXs (where TM = V, Cr, Mn, Fe, Co, Ni, Cu, Zn, and X = C, N, O, S, Se, Te). For each stable material we perform total energy mapping between the density functional theory corrected with on-site Coulomb repulsion (DFT+U) and the Heisenberg model Hamiltonian, for six particular magnetic orders - namely FM and AFM orders along three Cartesian directions (see Appendix B in \dag~Supplementary Information), in order to extract the governing magnetic interactions in the system. For the sake of screening, Hubbard parameter U in the calculations is taken from the online database, based on bulk oxides of transition metals \cite{onlineU}. In order to decrease computational cost, we consider only the first nearest-neighbor (NN) magnetic interactions and the single-ion anisotropy (SIA) in the Heisenberg Hamiltonian. 

The considered primitive unit cell of these materials consists of two TM, and two X atoms - such that each TM atom has four X atoms as the nearest neighbors, and vice versa (see Fig.~\ref{figure1}), where one can apply symmetry rules for the exchange matrix (see Appendix A in \dag~Supplementary Information) and SIA (cf. Ref.~\cite{Denis1}). In Fig.~\ref{figure1} each TM atom is sketched with one $d$ orbital, and each X atom with one $p$ orbital, since these atomic orbitals, and their mutual hybridization, are essential for the interactions between magnetic moments on TM atoms. Though the positions of the atoms in the structure are uniquely determined with respect to the in-plane primitive lattice vectors due to the symmetry of tetragonal structures, both atomic species are allowed to relax out-of-plane (along $\vec{a}_{3} \equiv z$ axis).

\begin{table}[ht]
\small
  \caption{\textbf{Magnetic properties of the stable monolayer tTMX structures.} FM, AFM, and NM stand for ferromagnetic, antiferromagnetic, and non-magnetic order, respectively. J${^{xx}}$ and J${^{yy}}$ mark in-plane, and J${^{zz}}$ out-of-plane exchange interactions. Out-of-plane exchange anisotropy $\Delta$ stands for the difference J${^{xx}}$-J${^{zz}}$. A${_{ii}^{zz}}$ is the single-ion anisotropy (SIA). If signs of SIA and exchange interactions are the same, SIA favors out-of-plane anisotropy, otherwise the in-plane one. J${^{xx}}$, J${^{yy}}$, J${^{zz}}$, $\Delta$, and A${_{ii}^{zz}}$ are all given in meV. T$_\textrm{c}$ stands for the critical temperature of the magnetic order, Curie for FM and N\'eel temperature for AFM monolayers. }
  \label{tblmagprop}
  \begin{tabular*}{0.5\textwidth}{@{\extracolsep{\fill}}lcccccc}
    \hline & Magnetic&\\&order & J${^{xx}}$=J${^{yy}}$ & J${^{zz}}$ & $\Delta~~$ &	A${_{ii}^{zz}}$ &	T$_\textrm{c}$(K) \\
    \hline
 CrC  &	 FM	 &	-52.68	 &	-52.66	&	0.03	&	-0.35	&	515.6	\\
 MnC  &	 FM	 &	-105.98 &	-106.01	&	-0.03   &	-0.67	&	1065.8	\\
 VN   &	 FM	 &	-3.73	&	-10.54	&	-6.81	&	-13.94	&	152.4	\\
 CoN  &	 FM	 &	-12.57	&	-12.94	&	-0.38	&	1.28	&	132.2	\\
 NiTe &	 FM	 &	-7.42	&	-5.15	&	2.27	&	4.53	&	31.2	\\
 FeC  &	 AFM &	53.00	&	53.28	&	0.28	&	0.21	&	5.3     \\
 FeO  &	 AFM &	5.47	&	5.46	&	-0.01	&	2.48	&	20.1	\\
 MnS  &	 AFM &	55.07	&	55.95	&	0.88	&	1.69	&	15.2	\\
 FeS  &	 AFM &	16.13	&	16.38	&	0.25	&	-0.56	&	5.7	    \\
 MnSe &	 AFM &	50.66	&	50.65	&	-0.01	&	-0.20	&	10.3	\\
 FeSe &	 AFM &	15.30	&	15.24	&	-0.06	&	-1.20	&	5.8	    \\
 FeTe &	 AFM &	31.80	&	-28.89	&	-60.69	&	119.63	&	35.6	\\
 CuC  &	 NM	 &	-	&	-	&	-	&	-	&	-	\\
 NiN  &	 NM	 &	-	&	-	&	-	&	-	&	-	\\
 CuN  &	 NM	 &	-	&	-	&	-	&	-	&	-	\\
 NiS  &	 NM	 &	-	&	-	&	-	&	-	&	-	\\
 CuS  &	 NM	 &	-	&	-	&	-	&	-	&	-	\\
 ZnS  &	 NM	 &	-	&	-	&	-	&	-	&	-	\\
 CuSe &	 NM	 &	-	&	-	&	-	&	-	&	-	\\
 ZnSe &	 NM	 &	-	&	-	&	-	&	-	&	-	\\
 ZnTe &	 NM	 &	-	&	-	&	-	&	-	&	-	\\
    \hline
  \end{tabular*}
\end{table}

Our throughput computational screening revealed that out of a 48 materials in total, only the 21 listed in Table~\ref{tblmagprop} possess dynamical stability. Out of those 21, we identified five materials with FM order as lowest in energy (out of six possibilities considered), seven materials with AFM order as the lowest-energy one, and nine materials not exhibiting magnetic order, i.e. with magnetic moments on individual atoms below 0.5$\mu_{B}$ in every of the six considered magnetic configurations, and the configuration with magnetic moments exactly 0 as lowest in energy. 
Out of five FM materials, two of them appear to have completely flat, square lattice when on-site Coulomb repulsion between electrons (U) is properly included - CrC and MnC. Such a structure ensures that the first-nearest-neighbor TM-X-TM bond comprises 90 degrees angle, and second-nearest-neighbor 180 degrees angle, making these structures optimal - according to GKA rules - for ``maximization" of magnetic exchange in the system. This fact served as motivation to thoroughly analyze the structural, electronic, and magnetic properties of these two materials, and discuss the subtle interplay of these properties that optimizes the ferromagnetic order.

\begin{figure}[t]
\centering
\includegraphics[width=\linewidth]{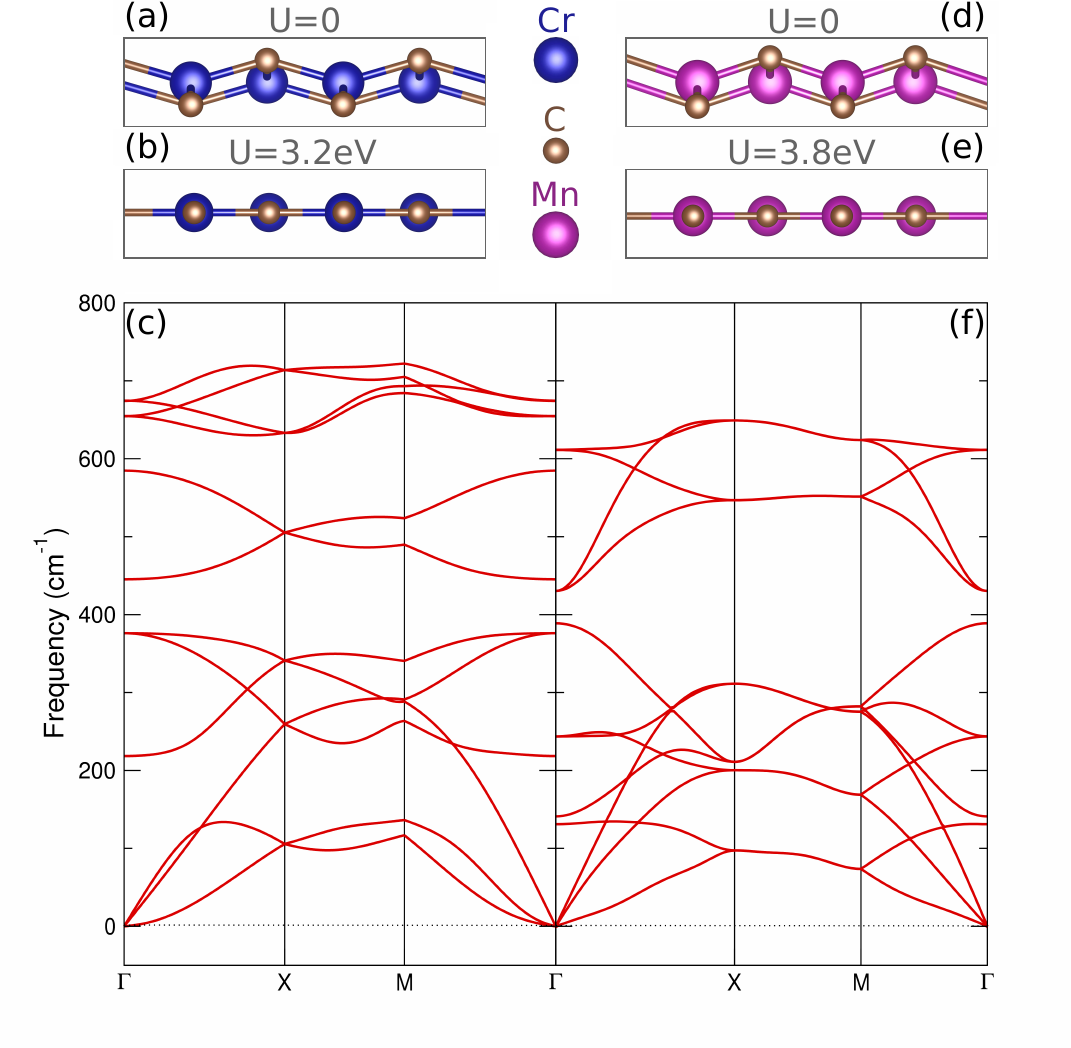}
\caption{\textbf{Stability of CrC and MnC ferromagnetic monolayers.} (a) The effect of Hubbard parameter U on monolayer structures of CrC and MnC, and (b) the phonon dispersions of two materials for optimal U shown in (a), proving dynamical stability of the flat structures.}
\label{fig:structandphonons}
\end{figure}

The first important finding concerns the structure. Namely, as indicated above, the relative atomic arrangement along the $z$ axis strongly depends on the on-site Coulomb repulsion (U) between electrons on TM atoms. In Fig.~\ref{fig:structandphonons}(a), we present the relaxed structures for CrC and MnC with no U included, and for the U value calculated self-consistently - using the linear response theory as introduced by Timorov \textit{et al.}~\cite{U-timorov-marzarri-cococionni-2022}. The latter U value is found as 3.2949~eV for CrC, and 3.8137~eV for MnC. 
In case of U=0, the structure buckles, and two X atoms relax above and below the plane of TM atoms. Furthermore, the X atoms create the tetrahedral structure around each TM atom. On the contrary, when using a realistic value for U, both CrC and MnC exhibit an ideally planar structure, i.e. a 2D square lattice as desired in the GKA argumentation towards the enhanced ferromagnetic order. For certainty, we tested the dynamical stability of both these planar structures - and the phonon dispersions shown in Fig.~\ref{fig:structandphonons}(b) exhibited no imaginary phonon frequencies.

After determining the on-site Coulomb repulsion and the planar structural stability, we move on to the magnetic properties of CrC and MnC. The necessary, yet insufficient condition for long-range magnetism is the non-zero magnetic moment per unit space. The TM atoms are expected to provide the latter, due to the localized unpaired electrons in their 3$d$ shell, each electron carrying spin $\frac{1}{2}$, and spin magnetic moment of $1~\mu_{B}$. In most of the structures based on TMs, the contribution of the orbital magnetic moment is negligible compared to the spin magnetic moment, due to the quenching of orbital momentum, hence the magnetic moment can be assumed to originate purely from the spin of the electron. Consequently, the total magnetic moment on each TM atom is $N\times1~\mu_{B}$, where $N$ is the total number of unpaired electrons in the 3$d$ shell of each TM atom. 

The basic ionic theory suggests $4+$ oxidation state of Cr and Mn cations in our monolayers, since C atom receives 4 electrons to reach stable octet configuration. Furthermore, Cr$^{4+}$ ion has 20 electrons and the $1s^{2}2s^{2}2p^{6}3s^{2}3p^{6}3d^{2}$ electronic configuration, while Mn$^{4+}$ ion 21 electrons and the $1s^{2}2s^{2}2p^{6}3s^{2}3p^{6}3d^{3}$ electronic configuration. This means that one expects $2~\mu_{B}$ per Cr atom and $4~\mu_{B}$ per primitive unit cell in CrC, and $3~\mu_{B}$ per Mn atom and $6~\mu_{B}$ per primitive unit cell in MnC. Ionic theory predicts no moment per C atom in either cases, due to the mentioned stable octet configuration.

For a more precise account of magnetic moments per atom we resort to DFT calculations, and find that: 
(1) in case of CrC, the magnetization per Cr atom is $2.8588~\mu_{B}$ and per C atom $-0.7236~\mu_{B}$, resulting in $4.2703~\mu_{B}$ per primitive unit cell;
(2) in case of MnC, the magnetization per Mn atom is $3.8825~\mu_{B}$ and per C atom $-0.7718~\mu_{B}$, resulting in $6.2214~\mu_{B}$ per primitive unit cell. The DFT results do corroborate the crude ionic theory regarding the total magnetization of the unit cell, but also reveal the significant hybridization between ($d$) orbitals of TM atoms and ($p$) orbitals of C atoms - causing a rather significant magnetization on otherwise non-magnetic C atoms.

However, in order for a system to host measurable magnetic order, next to the non-zero magnetic moments on TM ions, it must also host significant interaction between them. Unlike the initial estimates using the method based on mapping between total energies of the DFT and the Heisenberg Hamiltonians, we now calculate the magnetic exchange by mapping the energy variations due to the infinitesimal rotation of the magnetic moment on TM atoms from the reference FM state, between the DFT Hamiltonian in the localized-orbital basis set, and the Heisenberg Hamiltonian $H=\sum_{<i,j>} \textbf{S}{_i}\textbf{J}{_{i,j}}\textbf{S}{_j}$, as implemented in the TB2J code~\cite{TB2J2021}. In the latter Hamiltonian, $\textbf{S}_{i}$ denotes the unit 3D vector of the magnetic moment on the $i^{th}$ TM atom; $3\times3$ matrix $\textbf{J}_{i,j}$ stands for interaction between magnetic moments on $i$-th and $j$-th TM atoms; and $<i,j>$ denotes $i\neq j$, with avoided double counting. The SIA matrix cannot be calculated using this formalism (for details see \cite{TB2J2021}), therefore it is not explicitly written in the used Heisenberg Hamiltonian. However, SIA is not negligible, and correspondingly must be taken into account during e.g. the calculation of T$_\textrm{c}$ for the FM order. Therefore, we combine SIA reported in Table~\ref{tblmagprop}, together with the $\textbf{J}_{i,j}$ calculated using TB2J to construct the total model Hamiltonian for eventual 2$^{nd}$-principles calculations:
\begin{equation}\label{Heisenberg_Hamiltonian}
H=\sum_{<i,j>} \textbf{S}{_i}\textbf{J}{_{i,j}}\textbf{S}{_j} + \sum_{i} \textbf{S}{_i}\textbf{A}{_{i,i}}\textbf{S}{_i}.
\end{equation}

The main advantage of TB2J and the Green's-function-based methodology over the total energy mapping is the ability to orbitally disentangle the origins of  magnetic interactions~\cite{orb-res-CrI32020}, and also the ability to calculate interactions between \textit{all} different neighbors within a large supercell upon a \textit{single} DFT calculation on the primitive unit cell. In particular, we have calculated the matrices $\textbf{J}_{i,j}$ up to the 284$^{th}$ NN for both materials. 

\begin{figure}[t]
\centering
\includegraphics[width=\columnwidth]{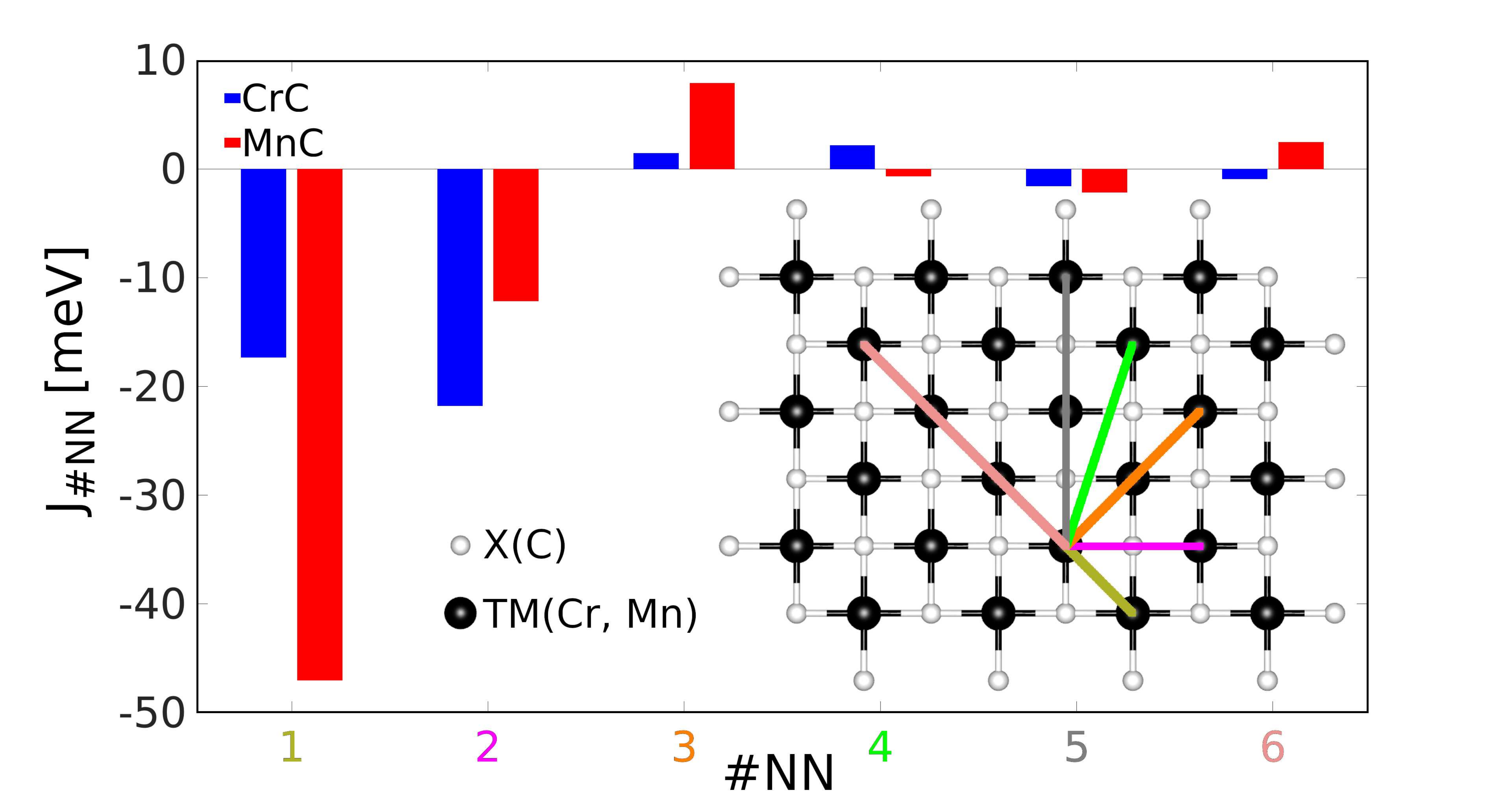}
\caption{\textbf{The isotropic magnetic exchange, per neighboring pair.} The isotropic exchange interactions in CrC (blue) and MnC (red), calculated using Green's function method as implemented in TB2J. The inset depicts the numerical labeling of the nearest-neighbor sites within the structure.}
\label{fig:JNN}
\end{figure}

Our results obtained using TB2J obey the symmetry-imposed constraints - i.e. all off-diagonal elements of all $\textbf{J}_{i,j}$ matrices are exactly 0. The diagonal part of each matrix can further be split - as presented in Table~\ref{tblmagprop} - into the isotropic (J$_{i,j} =$ J$^{xx}_{i,j} =$ J$^{yy}_{i,j}$) and the anisotropic part ($\Delta = $J$^{zz}_{i,j}-$J$^{xx}_{i,j}$).
The anisotropic part of the exchange, $\Delta$, in either system does not exceed few (1-6)~$\mu$eV, and is comparable with the rounding error in our calculations. Therefore, we consider $\Delta$ as effectively 0, and we subscribe the stabilization of the magnetic order in these 2D materials to just J$_{i,j}$ and SIA \footnote{This is in qualitative agreement with results reported in Table~\ref{tblmagprop}, where $\Delta$ was found to be much smaller than SIA and J$_{i,j}$.}. 

\begin{figure}
\centering
\includegraphics[width=\columnwidth]{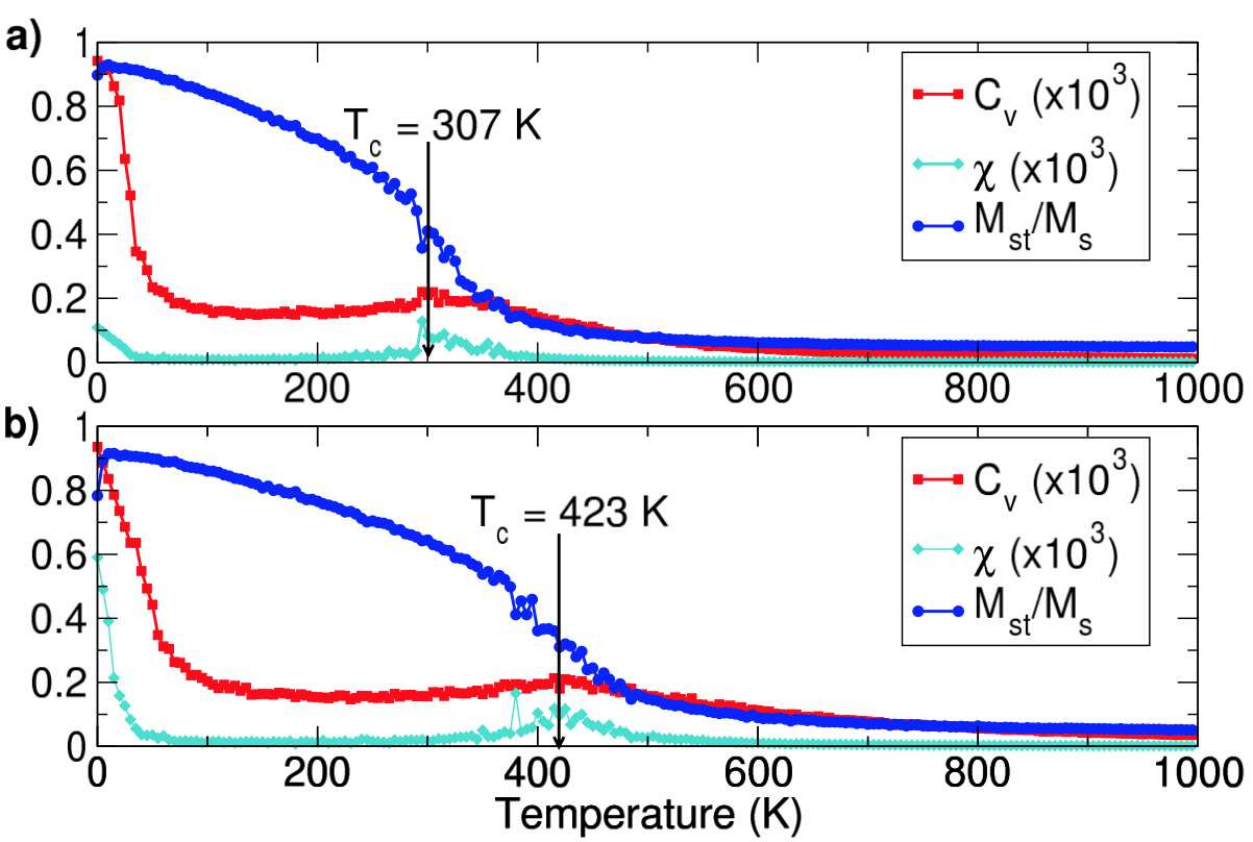}
\caption{\textbf{Thermal stability of the long-range magnetic order.} Magnetization, specific heat and magnetic susceptibility of (a) CrC and (b) MnC, as a function of temperature.}
\label{fig:temp}
\end{figure}

In Fig.~\ref{fig:JNN} we present our TB2J results for J$_{i,j}$ up to the 6$^{th}$ NN. One notices in Fig.~\ref{fig:JNN} that 1$^{st}$ and 2$^{nd}$ NN interactions are strong (few tens of meV), an order of magnitude larger  compared to usually encountered isotropic exchange values (few meV) in 2D materials. Another observation is that even pairs over 1~nm distance have small but non-zero exchange interaction, of few hundreds of $\mu$eV. However, since these are much smaller than the 1$^{st}$ and 2$^{nd}$ NN interactions, in what follows, we focus on the first two NN pairs with giant J$_{i,j}$, being essential for the high-T$_\textrm{c}$ GKA ferromagnetism. We use exchange interaction parameters calculated with TB2J, and SIA from Table~\ref{tblmagprop} to build the Heisenberg Hamiltonian as explained above, to then employ Monte Carlo simulations to evaluate the stability of the FM order with respect to temperature. In Fig.~\ref{fig:temp}, we present the thereby obtained evolution of (normalized) magnetization (M$_{st}$/M$_{s}$), magnetic susceptibility ($\chi$), and specific heat (C$_{v}$), as a function of temperature. One clearly sees that estimated critical temperatures of the FM order in both CrC and MnC \textit{exceed the room temperature}, being 307~K and 428~K respectively. Even though anisotropy is in general required to stabilize magnetic order in 2D above 0~K - in our case that is SIA - the main reason for such a large T$_\textrm{c}$ lies in the particularly large isotropic exchange between the 1$^{st}$ as well as the 2$^{nd}$ NN pairs of TM atoms. Table~\ref{tblmagprop} provides Curie and Neel temperature values for t-TMX structures. Notably, VN, CoN, and NiTe structures exhibit ferromagnetic properties; however, their respective Curie temperatures are significantly lower compared to those of CrC and MnC. This phenomenon can be primarily attributed to two key factors: weakened exchange interactions and strong Single Ion Anisotropy (SIA). This weakened exchange interaction and strong SIA inherently limits their ability to maintain ferromagnetic order at elevated temperatures, thus resulting in lower Curie temperatures when contrasted with CrC and MnC. Conversely, antiferromagnetic materials exhibit Neel temperatures in proximity to absolute zero (0 K) due to the significant impact of SIA. This low Neel temperature signifies the point at which antiferromagnetic materials undergo a transition to a non-magnetic state. The details about calculations of Curie and Neel temperatures can be found in \dag~Supplementary Information.

\section{Discussion} 

\begin{figure}
\centering
\includegraphics[width=\columnwidth]{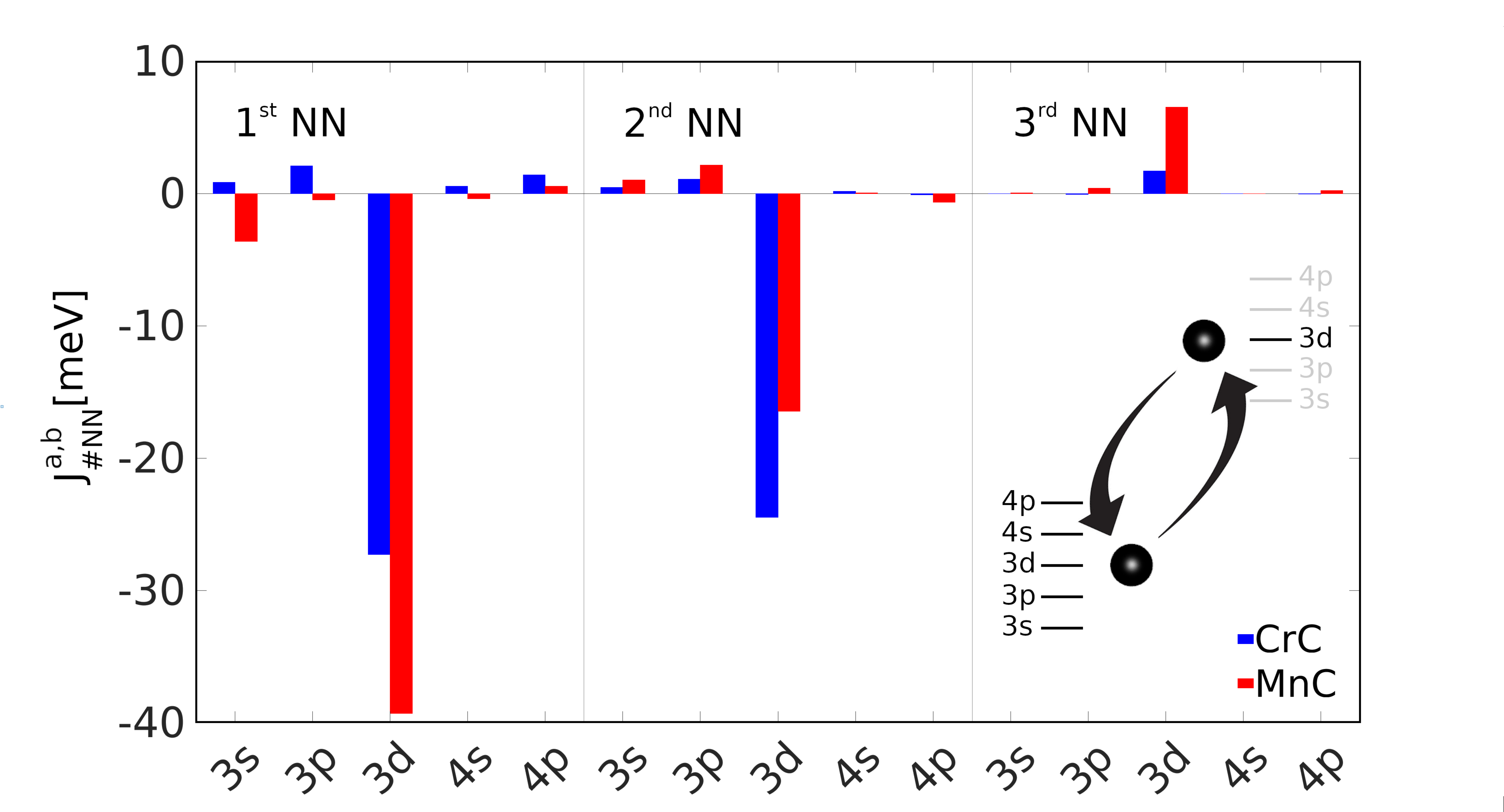}
\caption{\textbf{Orbitally-decomposed magnetic interactions.} The magnetic exchange interactions in CrC and MnC, for the three nearest-neighbor pairs. As illustrated in the inset, for every pair the contributions are discerned per orbital $a$ of TM1 ($a$ = $3s, 4s, 3p, 3d, 4p$), each interacting with just one orbital ($b$ = 3$d$) of TM2.}
\label{fig:orbital1}
\end{figure}

Having presented the core results, we next detail the origin of the magnetic interactions behind the observed high Curie temperature in monolayer tTMXs, bearing in mind the original assumptions following from the GKA rules. To shed the light on the source of the large exchange J$_{i,j}$, we look into the orbitally-resolved contributions.

\begin{figure*}
\centering
\includegraphics[width=\textwidth]{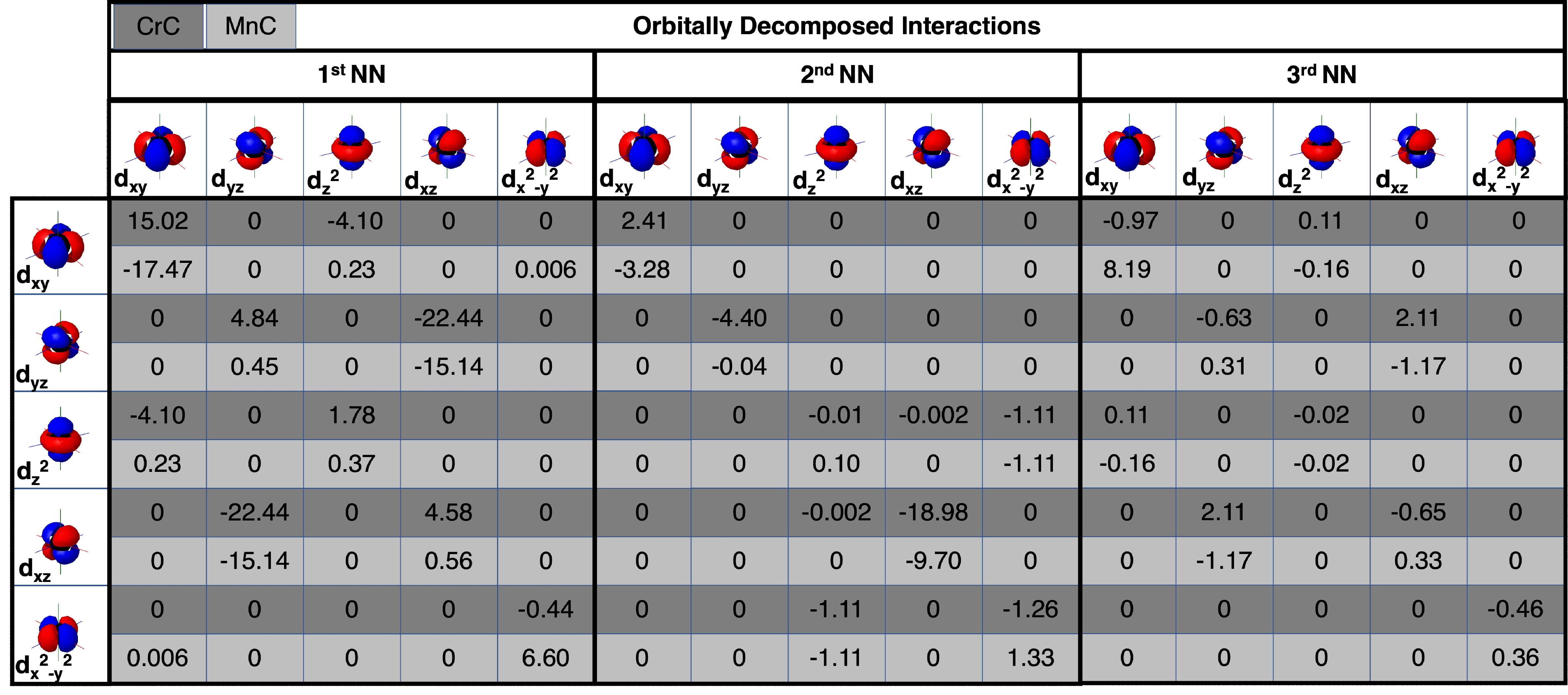}
\caption{\textbf{Sub-orbitally-decomposed $3d$-$3d$ interactions.} The magnetic exchange between five different 3d ($d_{xy}$, $d_{yz}$, $d_{z^2}$, $d_{xz}$ and $d_{x^2-y^2}$) orbitals on first three NN pairs, for both CrC and MnC monolayers. CrC and MnC are indicated using dark and light shading, respectively.}
\label{fig:orbital2}
\end{figure*}

Our initial assumption was that the gross of the exchange interactions originates from the exchange between the $d$ orbitals of the interacting TM atoms ($d$-$d$ exchange). In order to properly describe the physics behind our observations, we treat 3$s$, 3$p$, 3$d$, 4$s$, and 4$p$ as valence orbitals on TM atoms. Further, we quantify the contributions of each orbital-to-orbital interaction between the neighboring TM atoms (e.g. 4$s$ on TM1 and 3$d$ on TM2), to the total exchange between them. For facilitated interpretation, we consider the contributed interactions between each type of orbital on TM1 ($a$ = 3$s$, 4$s$, 3$p$, 3$d$, and 4$p$) and only 3$d$ orbitals on TM2 ($b$ = 3$d$), for the first three NN magnetic interactions $J^{a,b}_{\#NN}$, as shown in Fig. \ref{fig:orbital1}. It is rather obvious from Fig.~\ref{fig:orbital1} that TM1(3$d$)-TM2(3$d$) interactions dominate (being several tens of meV strong), and determine the magnetic order in the system - in this case the FM one. Interactions between other types of orbitals on one TM and 3$d$ orbitals on the other are generally at least an order of magnitude smaller. 
Therefore, after $d$-$d$ exchange is proven to be crucial for the large exchange and room-temperature magnetism in tTMXs, we next decompose it into the exchanges between individual $d$-orbitals of the interacting pair of TM atoms, to disentangle the key contributors. The results per $d$-orbital and per material are shown in Fig. \ref{fig:orbital2}.

\subsection{The nearest-neighbor interaction}
Owing to the square lattice symmetry of these monolayers, our original premise holds, and the empirical GKA rules are validated in the case of the 1$^{st}$ NN exchange - the ideally 90$^\circ$ TM-X-TM bond arrangement fosters a particularly strong FM interaction, stemming mainly from the $d$-$d$ exchange. 

As seen in Fig. \ref{fig:orbital2}, the largest contribution to the 1$^{st}$ NN magnetic interaction in both considered materials comes from the interaction between $d_{xz}$ on one TM atom, and $d_{yz}$ on the other, together with its symmetric twin - TM1($d_{yz}$)-TM2($d_{xz}$). We prescribe this exactly to the square-lattice geometry of the structure and the fact that the dumbbells of the $d_{xz}$ on TM1 and $d_{yz}$ on TM2 point along the bonds to the (X1) ligand atom between them in the structure (analogously for $d_{yz}$ on TM1 and $d_{xz}$ on TM2, interacting via the adjacent ligand X2). 
Moreover, after having a closer look at the DFT Hamiltonian in the localized basis set, we noticed that both $d_{xz}$ on TM1 and $d_{yz}$ on TM2 interact only with $p_{z}$ on X1, while the other hopping matrix elements (with $p_{x}$ and $p_{y}$ on X1, and $p_{x}$, $p_{y}$, and $p_{z}$ on X2) are 0, because of the symmetry of the system. 

Another significant contribution to the total exchange between the 1$^{st}$ NN pair originates from TM1($d_{xy}$)-TM2($d_{xy}$) interactions. In that case, the $d_{xy}$ orbitals on TM1 and TM2 point towards each other directly and have significant direct overlap, however, there is also significant hopping between $d_{xy}$ on both TM1 and TM2 and $p_{x}$ and $p_{y}$ orbitals on both nearest X atoms (X1 and X2). This increased complexity of the physical picture causes differences in the sign and the strength of those particular contributions in the two considered compounds: in case of CrC this contribution is AFM, while in case of MnC it is FM. The main reason for the difference in this orbital contribution lies in the different \textit{atomic environment} - the ordering and occupation of the atomic $d$($p$) orbitals on TM(X) atoms - and different behavior of direct exchange in the particular environment. In case of the atomic environment in CrC, the superexchange solely determines the orbital contribution to the total magnetic exchange and it is AFM, while in case of MnC, direct- and superexchange compete in such atomic environment that the result is the FM coupling. Details of these findings are made available in the \dag~Supplementary Information.

Both compounds host several other non-zero interactions between different $d$ orbitals, e.g. $d_{xy}$ on TM1 and $d_{z^2}$ on TM2. Even though these terms are much smaller than the dominating ones discussed above, they are still sizable (several meV) and they do affect the total exchange, albeit on tertiary level. In particular TM1($d_{xy}$)-TM2($d_{z^2}$) contribution is very sensitive to the alteration of the superexchange hopping TM($d_{xy}$)-X($p_{x/y}$), hence we note that superexchange mechanism plays an important role here. For more discussion on their origin and behavior, we refer the reader to the \dag~Supplementary Information.

\subsection{The next-nearest and further-neighbor interactions}
Although our initial premise of strong FM interactions between the first nearest neighbors was validated, we point out at this stage that plain GKA rules are not sufficient to interpret the magnetic behavior of a 2D material, even in the case of an ideally planar square-lattice structure. Namely, the 2$^{nd}$ NN interaction is expected to be AFM according to the GKA rules, due to the 180$^\circ$ TM-X-TM bond alignment - but we have observed the (strong) opposite in both materials of interest. 

In what follows, we present the results for the TM1-X-TM2 bonds being aligned with the global $x$ coordinate (as the case of TM1-X-TM2 along $y$ is completely analogous). As was the case with the 1$^{st}$ NN interaction, the orbitals aligned with the direction of TM-X-TM bonds are mainly responsible for the large total exchange of the 2$^{nd}$ NN pair as well - i.e. TM1($d_{xz}$)-TM2($d_{xz}$) interaction is the dominant one in this case. The dumbbells of these two $d$ orbitals both point towards the common X atom, and only interact with its $p_z$ orbital.  

The fact that the dominating contributions to the 1$^{st}$ and 2$^{nd}$ NN interactions are originating from the same physical process - i.e. from two $d$ orbitals on two TM atoms that are oriented towards the common ligand atom, and interact only with its $p_z$ orbital - leads towards the conclusion that they should be of the same sign and comparable strength, as they indeed are in our results (strongly FM). That said, the GKA rules assume the dominant contribution to the AFM superexchange in case of 180$^\circ$ TM-X-TM bonds to be via the $p$ orbital of X, whose dumbbell is aligned with TM-X-TM direction~\cite{vanVleck,kanamori} - which would be the $p_x$ orbital in the above discussion. However, even though we find these contributions to be AFM as GKA rules would suggest, we also find that they are an order of magnitude smaller than the dominant contributions, involving the $p_z$ orbital. Therefore, the disagreement between our results and GKA rules originates in the fact that mechanism considered dominant by GKA (superexchange involving the $p_x$, or $p_\sigma$) is secondary in our case, and vice versa - the mechanism considered secondary by GKA (superexchange involving the $p_z$, or $p_\pi$) appears to be dominant in the two materials of our interest. For interested readers, this is discussed in more detail within the \dag~Supplementary Information.

In case of the 3$^{rd}$ NN exchange, the geometry is very similar to the 1$^{st}$ NN exchange, hence one expects to have the same dominant contributions. We find that the three main contributions in both materials are indeed the same ones as in the case of the 1$^{st}$ NN, i.e. $d_{xz}$-$d_{yz}$, $d_{yz}$-$d_{xz}$, and $d_{xy}$-$d_{xy}$. However, their observed behavior is more complicated. In case of CrC the TM1($d_{xz}$)-TM2($d_{yz}$) and TM1($d_{yz}$)-TM2($d_{xz}$) are contributing the most, however to AFM order (positive exchange parameter), while the second largest contribution comes from TM1($d_{xy}$)-TM2($d_{xy}$) and it is FM. 
In case of MnC the main contribution comes from TM1($d_{xy}$)-TM2($d_{xy}$) and it is AFM. The interaction between TM1($d_{xz}$)-TM2($d_{yz}$) and TM1($d_{yz}$)-TM2($d_{xz}$) is again FM, however their magnitude affects the total 3$^{rd}$ NN exchange significantly less. 
Since the 3$^{rd}$ NN exchange is negligible compared to the 1$^{st}$ and 2$^{nd}$ NN exchange, we will not detail these interactions. We see however that sign and strength of the orbital contributions to the 3$^{rd}$ NN exchange are mainly determined by the orbital ordering and occupation of Cr and Mn atoms - i.e. their atomic environment - and the fact that in different environment, the different mechanisms may be dominating. In case of the TM1($d_{xy}$)-TM2($d_{xy}$) contribution, our results suggest that in case of CrC there is the competition between direct and superexchange, while in MnC it is the usual superexchange through the X ligand that dominates. By comparing the results for the first- and the third-nearest-neighbor exchange in two materials, one could argue that in case that atomic environment stimulates the superexchange alone, the result will be AFM interaction (the first-nearest-neighbor in CrC, and third-nearest-neighbor in MnC). On the other hand, when atomic environment stimulates the competition between different exchange mechanisms (the first-nearest-neighbor in MnC, and third-nearest-neighbor in CrC) resulting interaction between TM1($d_{xy}$)-TM2($d_{xy}$) will be FM. For interested readers, we provide brief additional discussion of the effect of atomic environment to the third-nearest-neighbor magnetic exchange in the \dag~Supplementary Information.

\section{Conclusions}
In summary, after a computational throughput screening of the whole family of tTMX materials, we have identified two dynamically stable and ideally 2D flat Lieb-like magnetic crystals, CrC and MnC. According to the seminal Goodenough-Kanamori-Anderson rules, materials of such symmetry are prone to host pronouncedly high ferromagnetic exchange interaction. Our detailed analysis of the magnetic properties has shown that both Cr and Mn ions indeed have sizable magnetic moments, that materials host large isotropic magnetic interactions (order of 10 meV), that exchange anisotropy is negligible in these systems (order of 1 $\mu$eV), and that stabilization of the long-range magnetic order in these materials should be prescribed to single-ion anisotropy (order of 0.1 meV). As a result of this large isotropic exchange and non-zero single-ion anisotropy, the Curie temperature for ferromagnetic transition of these materials exceeds the room temperature - 307 K in CrC and 428 K in MnC.

In detailed analysis, we showed how the symmetry/geometry of the system selects the dominant contributions to the total magnetic exchange. In case of the 1$^{st}$ NN and 90 degrees TM-C-TM lattice direction, the interaction between $d_{xz}$ on TM1 and $d_{yz}$ on TM2 - TM1($d_{xz}$)-TM2($d_{yz}$) - dominates together with its twin counterpart TM1($d_{yz}$)-TM2($d_{xz}$). We presented the hypothesis that this interaction is mediated by the $p_z$ orbital of the intervening ligand C. In case of the 2$^{nd}$ NN and 180 degrees TM-C-TM lattice direction, the dominant contributions are again geometry-selected, being either TM1($d_{xz}$)-TM2($d_{xz}$) or TM1($d_{yz}$)-TM2($d_{yz}$), depending on whether TM-C-TM direction is along global Cartesian $x$ or $y$ coordinate. In this case we also hypothesized that interaction is mediated by the $p_z$ orbital of the C atom.
In case of the 3$^{rd}$ NN, which was of secondary importance in our systems, we could not establish a criterion for single dominant contribution based on geometry, however, even there two contributions were identified as leading ones. Finally we have shown that interactions beyond the 3$^{rd}$ NN are far smaller and may be neglected in any prediction of experimentally measurable physical properties of these systems.

In conclusion, we provided convincing evidence that CrC and MnC in their two-dimensional tetragonal phase are room-temperature Lieb-like-lattice 2D ferromagnets. Furthermore, we highlighted the effect of geometry on the magnetic properties - both the magnetic moments and the exchange interactions between them. That said, we expect any emergent material with flat square-lattice structure and magnetic atom from the first half of the first row of transition metals (V, Cr, Mn, Fe) to: (i) have sizable magnetic moments (mainly from unpaired $d_{xz}$ and $d_{yz}$ orbitals), and (ii) host large geometry-selected magnetic interactions between these two types of orbitals. Consequently, in any such system with sufficient anisotropy in magnetic interactions, the latter features will foster the large isotropic exchange and high critical temperature of the long-range magnetic order in two dimensions, as already shown for 2D metallic MnB monolayer.\cite{MnB}

\section{Acknowledgement}
This work was supported by the Research Foundation-Flanders (FWO) and the Technological Research Council of Turkey (TUBITAK) under Contract No. 118F512. D.Š. is a doctoral fellow of FWO under Contract No. 11J4322N. The computational resources and services for this work were provided by the VSC (Flemish Supercomputer Center), funded by the FWO and the Flemish Government -- department EWI.

\bibliography{ref}

\begin{mcitethebibliography}{41}
\bibitem[Note()]{Note}
This is in qualitative agreement with results reported in Table~\ref {tblmagprop}, where $\Delta $ was found to be much smaller than SIA and J$_{i,j}$.\relax
\mciteBstWouldAddEndPunctfalse
\mciteSetBstMidEndSepPunct{\mcitedefaultmidpunct}
{}{\mcitedefaultseppunct}\relax
\EndOfBibitem
\end{mcitethebibliography}
\bibliographystyle{ref}
\ifarXiv
    

\section*{Supplementary material}

# Goodenough-Kanamori-Anderson high-temperature ferromagnetism in tetragonal transition-metal xenes

U. Yorulmaz,[1, 2] D. Šabani,[2] C. Sevik,[3, 2] and M. V. Milošević[2]

[1]*Department of Physics, Eskisehir Osmangazi University, Eskisehir, Turkiye*
[2]*Department of Physics & NANOlab Center of Excellence,*
*University of Antwerp, Groenenborgerlaan 171, B-2020 Antwerp, Belgium*
[3]*Department of Mechanical Engineering, Eskisehir Technical University, Eskisehir, Turkiye*

## I. DETAILS OF COMPUTATIONAL CALCULATIONS

We performed calculations in the framework of density functional theory (DFT)[1] using linear combination of atomic orbitals (LCAO) as implemented in the SIESTA code [2]. And also a planewave based codes VASP [3–5] and QUANTUM ESPRESSO [6] were used. In VASP, the generalized gradient approximation (GGA) were treated with the Perdew–Burke–Ernzerhof (PBE) form for the effect of electron exchange–correlation [7]. In SIESTA and QUANTUM ESPRESSO, PBE norm-conserving, fully relativistic pseudopotentials in the GGA which were generated by the PseudoDojo project were used [8]. The electronic density converges with an energy cutoff of 600 eV and Monkhorst–Pack k-grid of $12 \times 12 \times 1$. To prevent interactions between periodic layers along the z-direction, a slab model with a least 15 Å vacuum thickness was used. The convergence criteria for the aim of achieving a high level of accuracy were chosen the total energy lower than $10^{-7}$ eV and the force acting on each atom was below $10^{-7}$ eV$Å^{-1}$. The interatomic magnetic interactions are calculated by Green's function method based on magnetic force theory (MFT) as implemented in TB2J code [9]. Since this code uses Wannier functions or linear combinations of atomic orbitals, Hamiltonians obtained with VASP and QE cannot be used directly in the TB2J code. Therefore, Wannier90 code was used to calculate Maximally localized Wannier functions (MLWFs) [10]. Finally, the effect of temperature on magnetic properties was investigated by using a second-principles approach for lattice dynamics simulations, implemented as the MULTIBINIT code in the ABINIT code package [11].

## II. ORBITALLY-RESOLVED ISOTROPIC MAGNETIC EXCHANGE

The state of the art in describing the origin of magnetic interactions in a system is to resolve the magnetic interactions on orbital level - as we presented in Fig. 7 in the main text - where one decomposes the total exchange between two magnetic atoms into a $5 \times 5$ matrix whose each element corresponds to the magnetic exchange between certain $d$ orbital on atom A and another $d$ orbital on atom B. However in general, exchange between any two orbitals on adjacent magnetic (M) atoms contains both contributions from direct exchange between $d$ orbitals and superexchange between $d$ orbitals through the $p$ orbitals of non-magnetic (NM) atoms. Furthermore, these contributions are mixed, and at this point, even the most advanced theoretical framework [12, 13] is unable to separate them.

However, we introduce the approach that allows one to quantify the influence of direct- and super-exchange on different orbitally-resolved magnetic parameters. The core of the idea is that after obtaining DFT Hamiltonian in tight-binding (TB) basis of localized single-particle orbitals for a certain material, one can sagely tune the particular TB hopping parameters connected with the direct exchange (M1($d$)-M2($d$) hopping parameters) and superexchange (M($d$)-NM($p$) hopping parameters) and quantify the influence on different orbital-level magnetic interactions.

Before reporting our findings using this approach to the systems of our interest, we must comment on a subtle yet important detail to practically execute the idea. The results reported in the main text were obtained by using the double-valence-single-polarization-$\zeta$ localized basis or atomic orbitals (SIESTA DFT code [14–16]) while the analysis conducted in this document is based on results obtained by projection of DFT results (VASP [3–5]) to the maximally-localized Wannier functions (MLWFs) basis [10].

SIESTA was the obvious choice to calculate orbitally-resolved magnetic interactions mainly due to the fact that Hamiltonian was intrinsically in the localized basis, with no need to perform additional projections from plane-wave-like to the basis of localized orbitals. However, problems appear if one desires to associate a physical meaning to the hopping matrix elements and to tune them. Due to the fact that occupied atomic orbitals are split into two (double-valence-$\zeta$), there is no single hopping element, but rather four of them representing the interaction between those to occupied atomic orbitals. This adds certain difficulty in the core of the idea - to tune the hopping parameter and see

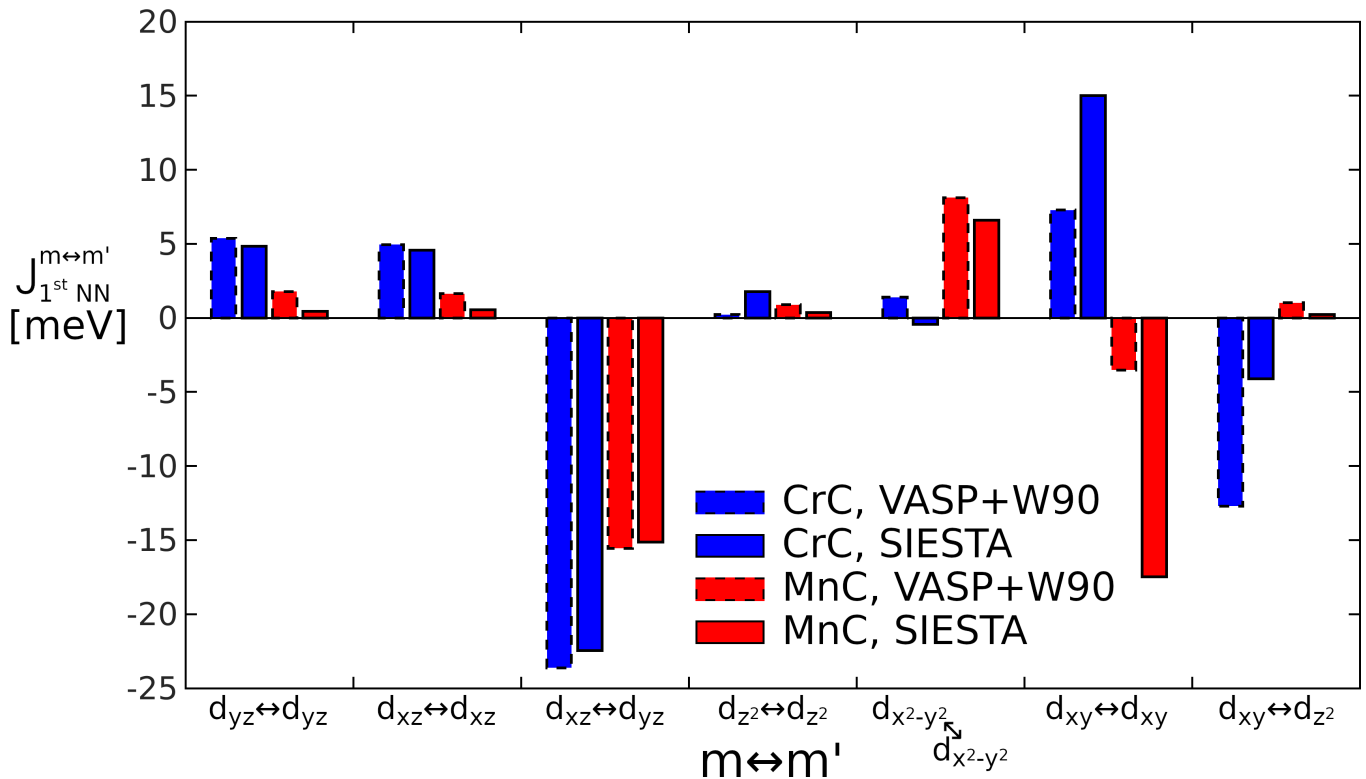

FIG. 1: Comparison of the results for the first-nearest-neighbor orbitally-resolved magnetic exchange in two materials - CrC and MnC, using the two computational choices - SIESTA and VASP+Wannier90. Only seven non-zero contributions are presented.

the outcome - since it is unclear what and how exactly should be tuned. Furthermore, even though we accomplished to read the binary data from an output file representing the Hamiltonian operator in the SIESTA localized basis, another issue - next to the mentioned tuning problem - was writing back the data to the original binary format of the .HSX file. All these reasons together guided us to choose MLWFs where single MLWF is assigned to the single atomic orbital and Hamiltonian in localized basis is written in human-readable format. To convince the reader that two choices are near equivalent, we compare the results for the first-nearest-neighbor orbitally-resolved magnetic exchange in the Fig. 1. From Fig. 1 one can see that all but one contributions are qualitatively the same, the only difference being present in the interaction between two $d_{x^2-y^2}$ orbitals on adjacent Cr atoms in CrC, one of the least important contributions to the total magnetic exchange in that material. Regarding the quantitative differences, they are mainly present in (1) M1($d_{xy}$)-M2($d_{xy}$) and (2) M1($d_{xy}$)-M2($d_{z^2}$). Moreover, the same two contributions have different signs in two materials, regardless of the computational setup used, and due to everything said they will be in the focus of further discussion, although the stronger contributions are present in both systems, i.e. M1($d_{xz}$)-M2($d_{yz}$) and its twin M1($d_{yz}$)-M2($d_{xz}$).

## A. Tuning the direct- and super-exchange in DFT Hamiltonians

We commence the analysis with varying two sorts of hopping elements between zero and their nominal value in each material. The first are the Hamiltonian elements connected with hopping between $d_{xy}$ orbitals on adjacent atoms - associated with the direct exchange - and the second are the Hamiltonian elements between $d_{xy}$ orbitals on TM atom and $p_x$ or $p_y$ on neighboring ligand - associated with the superexchange. We present the effect of the variation of those Hamiltonian hopping elements on the seven non-zero orbitally-resolved contributions to the total first-nearest-neighbor exchange in the Figs. 2 and 3 for CrC and MnC, respectively.

The results presented in latter figures suggest several remarks to make. The first remark is that variation of $d_{xy}$-$d_{xy}$ and $d_{xy}$-$p_{x/y}$ Hamiltonian matrix elements had no effect on $J_{d_{xz}-d_{yz}}$, $J_{d_{xz}-d_{xz}}$, and $J_{d_{yz}-d_{yz}}$. This gives an important hint that electronic states are divided into independent classes with no entanglement in-between. Second remark is

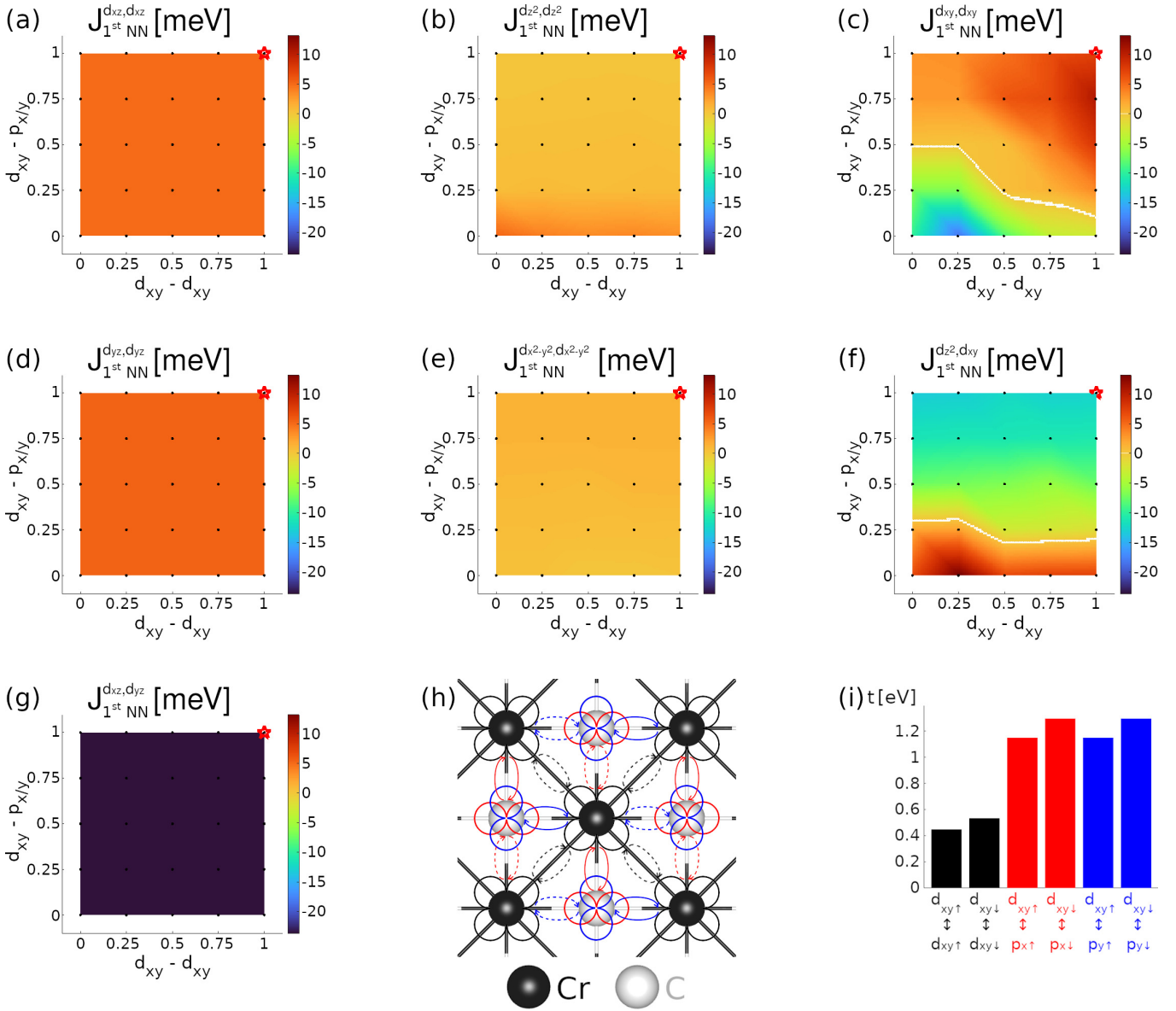

FIG. 2: The sensitivity of the first-nearest-neighbors magnetic exchange w.r.t. orbital interactions connected with direct- and super-exchange in case of CrC. (a)-(g) The seven non-zero orbital contributions to the total exchange. (h) The sketch of relevant orbitals in the system of interest, black orbital is $d_{xy}$ on Cr atoms, while red and blue orbitals are $p_x$ and $p_y$ from C atoms. The arrows mark interactions between orbitals: black arrows corresponds to $d_{xy}$-$d_{xy}$, red arrows to $d_{xy}$-$p_x$, and blue arrows to $d_{xy}$-$p_y$; solid-line arrows represent positive hopping between two orbitals, while dashed-line arrows stand for negative hopping parameter. The magnitudes of the three mentioned hopping parameters, for both up and down spins, are illustrated in (i).

that the sign of the $J_{d_{xy}-d_{xy}}$ in case of both direct- and super-exchange included [(1,1) point in panel (c), marked with the red star] is the same as the sign when only superexchange is included [(0,1) point in (c)] and is different than the sign when only direct exchange is included [(1,0) point in (c)]. This suggests that even though this contribution has different sign in two materials, in both cases the superexchange determines its sign - in case of CrC, it works for AFM order and in case of MnC for FM alignment. The third and final remark is that the "path of changing sign" - white line in Figs. 2(c) and 3(c) - is completely different in two materials, and consequently the way how direct- and super-exchange compete is different in two materials. Since the hopping matrix elements in two materials are quantitatively close [Figs. 2(i) and 3(i)], one hypothesis to be tested is that atomic *environment* plays important role in determining the influence of direct- and super-exchange to the orbitally-resolved contributions. A certain

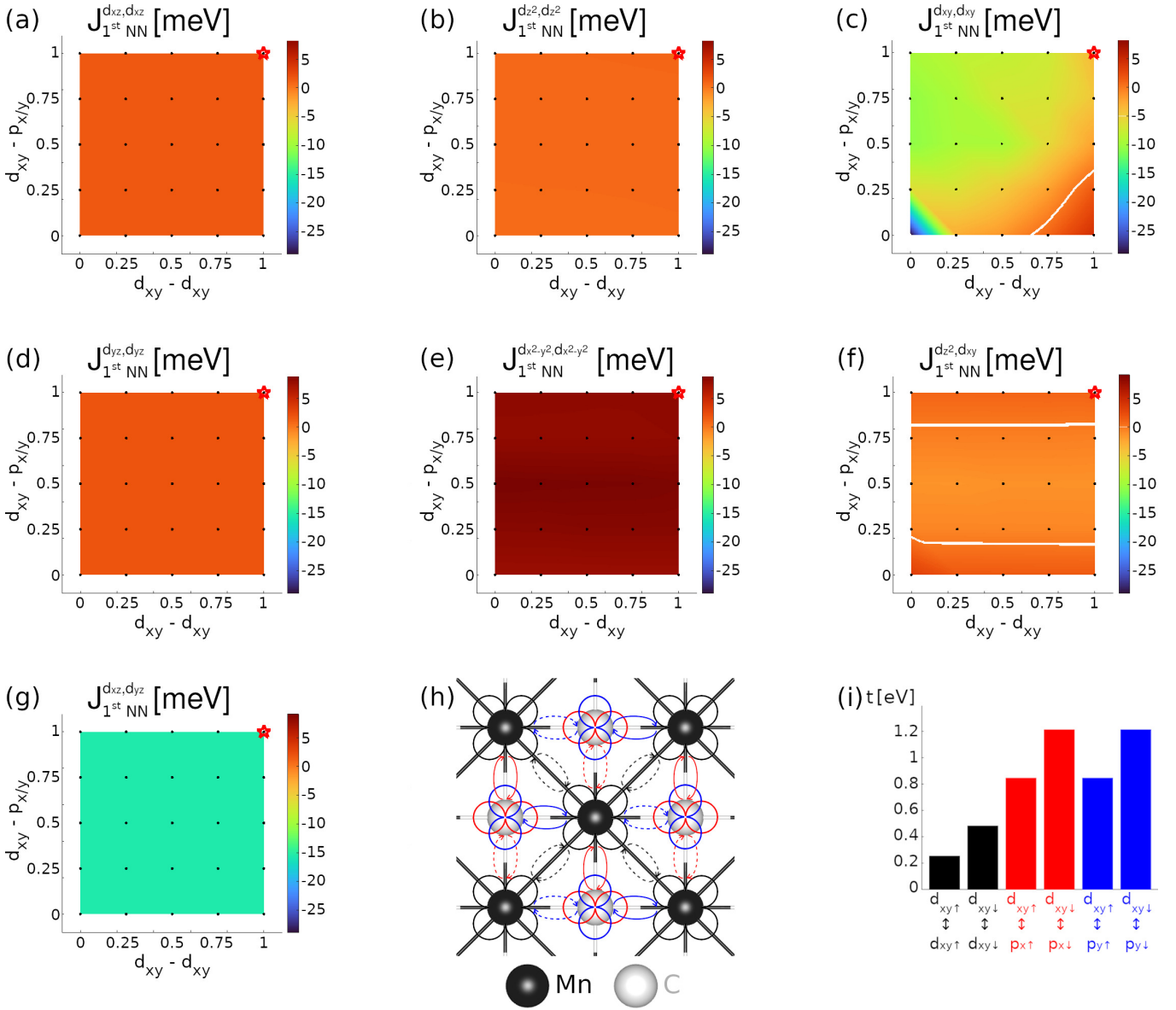

FIG. 3: The sensitivity of the first-nearest-neighbors magnetic exchange w.r.t. orbital interactions connected with direct- and super-exchange in case of MnC. (a)-(g) the seven non-zero orbital contributions to the total exchange. (h) The sketch of relevant orbitals in the system of interest, black orbital is $d_{xy}$ on Cr atoms, while red and blue orbitals are $p_x$ and $p_y$ from C atoms. The arrows mark interactions between orbitals: black arrows corresponds to $d_{xy}$-$d_{xy}$, red arrows to $d_{xy}$-$p_x$, and blue arrows to $d_{xy}$-$p_y$; full-line arrows represent positive hopping between two orbitals, while dashed-line arrows stand for negative hopping parameter. The magnitudes of three mentioned hopping parameters, for both up and down spins are illustrated in (i).

environment is created in the system by electrons occupying certain localized ($d$ or $p$) states and leaving others empty. To illustrate the most obvious difference in the two atomic environments - in case of CrC three $d$ orbitals ($d_{xz}$, $d_{yz}$, and $d_{z^2}$) with spin up are occupied on Cr, and in case of MnC all five $d$ orbitals with spin up are occupied on Mn. We will further analyze the three remarks from above in detail, in the same order.

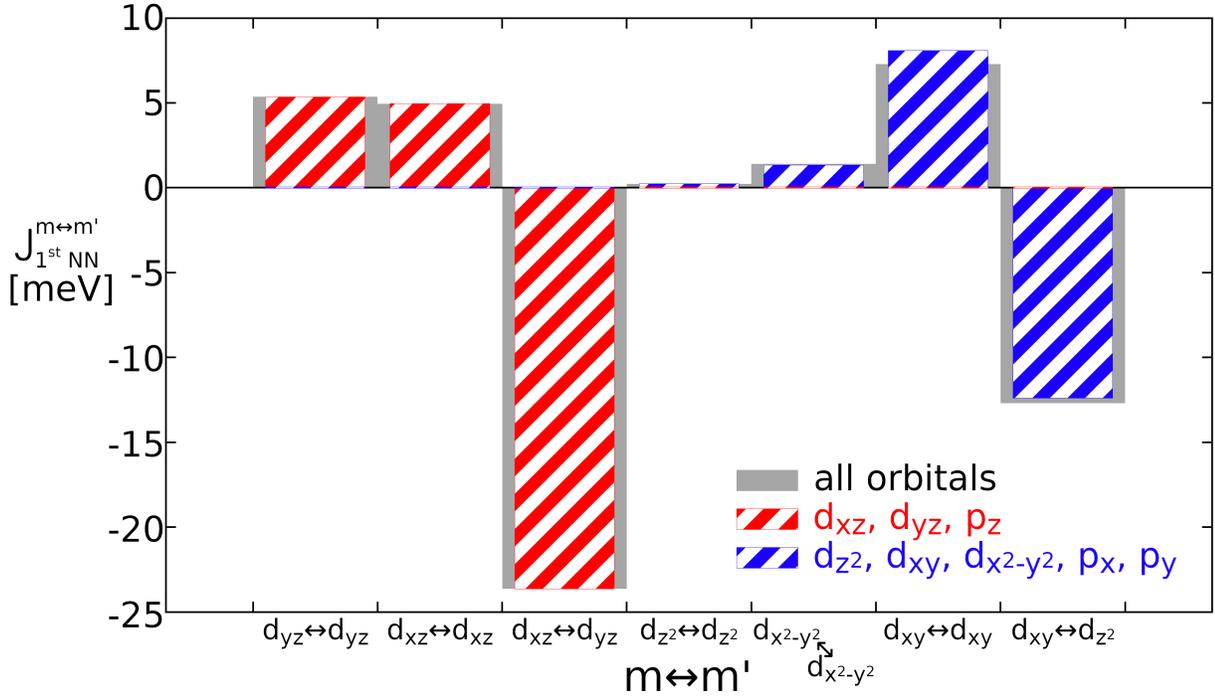

(a) Choice #1: class A - $d_{xz}$, $d_{yz}$ and $p_z$; class B - $d_{z^2}$, $d_{xy}$, $d_{x^2-y^2}$ and $p_x$, $p_y$.

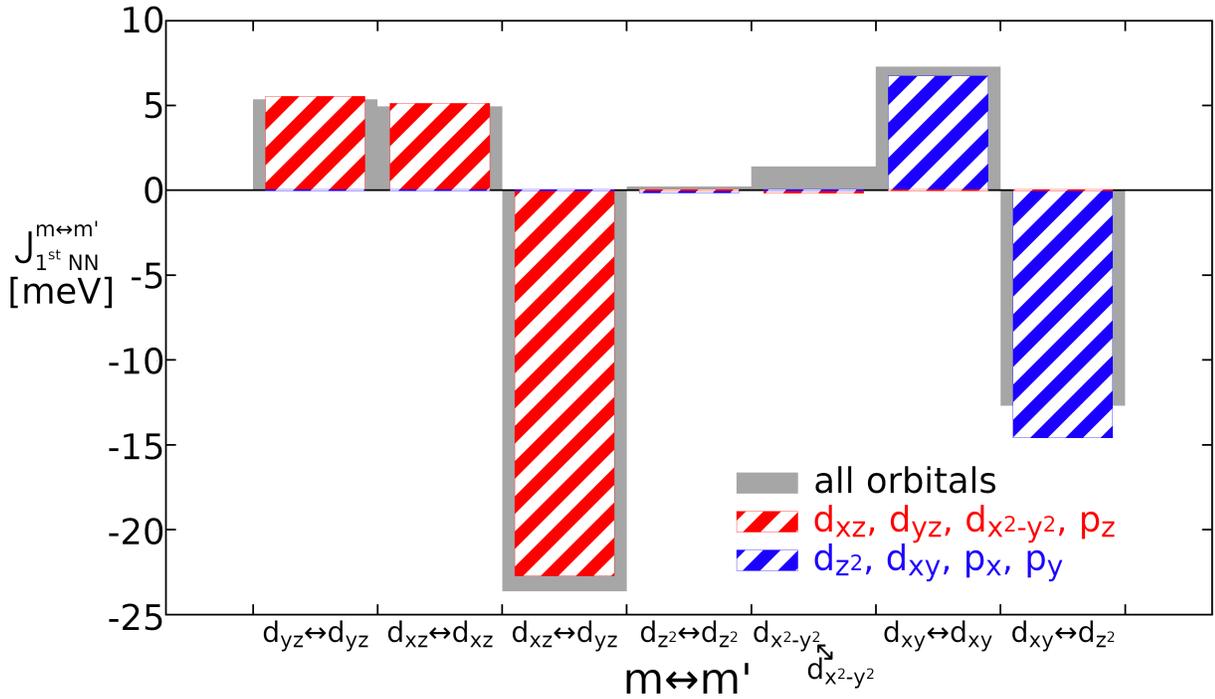

(b) Choice #2: class A - $d_{xz}$, $d_{yz}$, $d_{x^2-y^2}$ and $p_z$; class B - $d_{z^2}$, $d_{xy}$ and $p_x$, $p_y$.

FIG. 4: Proof for the separation of non-interacting classes A and B of orbitals in case of the first-nearest-neighbor interactions in CrC. Grey area on the graphs is obtained when all orbitals are taken from the original CrC Hamiltonian. Red-white bars are obtained when only class A is taken from CrC Hamiltonian and class B elements are set to 0. Blue-white bars are obtained when only class B is taken from CrC Hamiltonian and class A elements are set to 0.

## B. Separation of localized orbitals into two classes

To confirm the behavioral separation of the electronic states into two classes, we focus on the first-nearest-neighbor interactions in CrC. We group electronic states - i.e. $d$ orbitals from Cr atoms and $p$ orbitals from C atoms - into two classes: class A with $d_{xz}$, $d_{yz}$, and $p_z$; and class B with $d_{z^2}$, $d_{x^2-y^2}$, $d_{xy}$, $p_x$ and $p_y$. Moreover, we perform three sets of calculations: the first set that corresponds to the CrC, where we include both classes, the second set where we include only class A, while matrix elements dealing with orbitals from class B are set to 0, and the third set where we include only class B, while matrix elements connected with class A are zeroed. The results of this analysis are presented in the Fig. 4a. These results confirm our initial expectation that electronic states of two classes are practically independent and non-interacting from the point of view of magnetic exchange. One can see that when only states from class A are included and exchange between orbitals from class B is zero, the exchange between those $d$ orbitals from class A remains unchanged compared to the full CrC case, and vice versa.

However, due to the fact that the main differences between two materials (and between two DFT codes used on same material) originate from $J_{d_{xy}-d_{xy}}$ and $J_{d_{xy}-d_{z^2}}$, we tested the idea of shifting $d_{x^2-y^2}$ from class B to class A, and leaving only $d_{xy}$ and $d_{z^2}$ with $p_x$ and $p_y$ in class B. The results of this modified division are presented in Fig. 4b. One can notice that the shift, as expected, affected one of the smallest contributions, $J_{d_{x^2-y^2}-d_{x^2-y^2}}$ drastically, while other, more dominant contributions changed only slightly. In any case, we have shown that the core of the class A contains $d_{xz}$, $d_{yz}$, and $p_z$; the core of the class B contains $d_{z^2}$, $d_{xy}$, $p_x$ and $p_y$; and the $d_{x^2-y^2}$ can be sorted into any of the two classes, without much effect on the magnetism in the system.

To complete the study about the separation of electronic states in two materials we have performed two additional calculations for both the first- and the second-nearest-neighbor magnetic interactions in each material. Next to the regular calculations where both class A and B are taken from CrC or MnC, we performed two "mixed" calculations: one where class A is taken from CrC and class B from MnC, and another where class A is taken from MnC and class B from CrC. The results of the regular calculations (already shown in Fig. 1) are presented together with the results of the two "mixed" calculations in Figs. 5a and 5b. These results provide additional proof that classes A and B are not entangled whatsoever. One can deliberately alter them and only those contributions to magnetic exchange containing $d$ orbitals from the corresponding class will be affected. In Figs. 5a and 5b one can see that green bar follows blue bar only when $d$ orbitals from the class A are involved - $d_{xz}$, $d_{yz}$, and $d_{x^2-y^2}$. The same green bar follows the red bar in case of $d$ orbitals from the class B - $d_{xy}$, $d_{z^2}$. The situation with the yellow bar is exactly opposite - it follows the red bar for $d$ orbitals from the class A and blue bar for $d$ orbitals from the class B. This is consistent for all contributions, in case of the first- and the second-nearest-neighbor magnetic exchange.

## C. The second-nearest-neighbor magnetic exchange - challenging the GKA rules

Before we continue with the competition between direct- and super-exchange, and the effect of the atomic environment and orbital occupation, here we will comment on the results for the second-nearest-neighbor exchange. Out of 25 contributions to the total exchange (in general 5×5 matrix), only 6 of them are non-zero. Unlike one would guess based on the Goodenough-Kanamori-Anderson (GKA) rules, here the total second-nearest-neighbor magnetic exchange is FM in both materials. The main contribution comes from the geometry-selected $J_{d_{xz}-d_{xz}}$, since two second-nearest-neighbor M atoms lie on $x$ axis. Equivalently, for two second-nearest-neighbor M atoms on $y$ axis, $J_{d_{yz}-d_{yz}}$ is dominant. These dominant terms in second-nearest-neighbor magnetic interactions are equivalent to the dominant term in the first-nearest-neighbor magnetic exchange, in sense that shape of orbitals participating in the interaction follows the M-M bridge through the shared carbon ligand. Therefore it is not surprising that the sign of the first- and the second-nearest-neighbor magnetic exchange is the same, and that their strengths are comparable.

Contrary to our results, GKA rules would predict that the second-nearest-neighbor interaction is AFM, however, it is important to mention that conclusions of GKA rules are derived in the octahedral surrounding of the M atom, while in our case the surrounding is square-planar, which considerably affects the splitting and occupation of $d$ orbitals on M atoms. In addition, in case of GKA, the total magnetic exchange is assumed to mainly originate from the superexchange through the $p_\sigma$ (see Section 1, Section 2, and Fig. 1 in Ref. [17], or sketch in Section 5 of Ref. [18]). In the particular case of M-M direction in our calculation - the $x$ axis - $p_\sigma$ corresponds to $p_x$ orbitals. Despite different atomic environments, our results confirm that two orbitally resolved contributions where $p_x$ (or $p_\sigma$) is playing a role are both AFM - $J_{d_{z^2}-d_{z^2}}$ and $J_{d_{x^2-y^2}-d_{x^2-y^2}}$ - nevertheless, they are both very weak.

Additionally, GKA rules consider another process which contributes to the total superexchange - the process involving one of the two $p_\pi$ orbitals (in our case this orbital is $p_z$, see also Fig. 2 in Ref. [17]). In case of GKA rules, this mechanism is considered as secondary because, to quote Kanamori: "*this $d_\epsilon - p_\pi$ bond should be weaker than the*

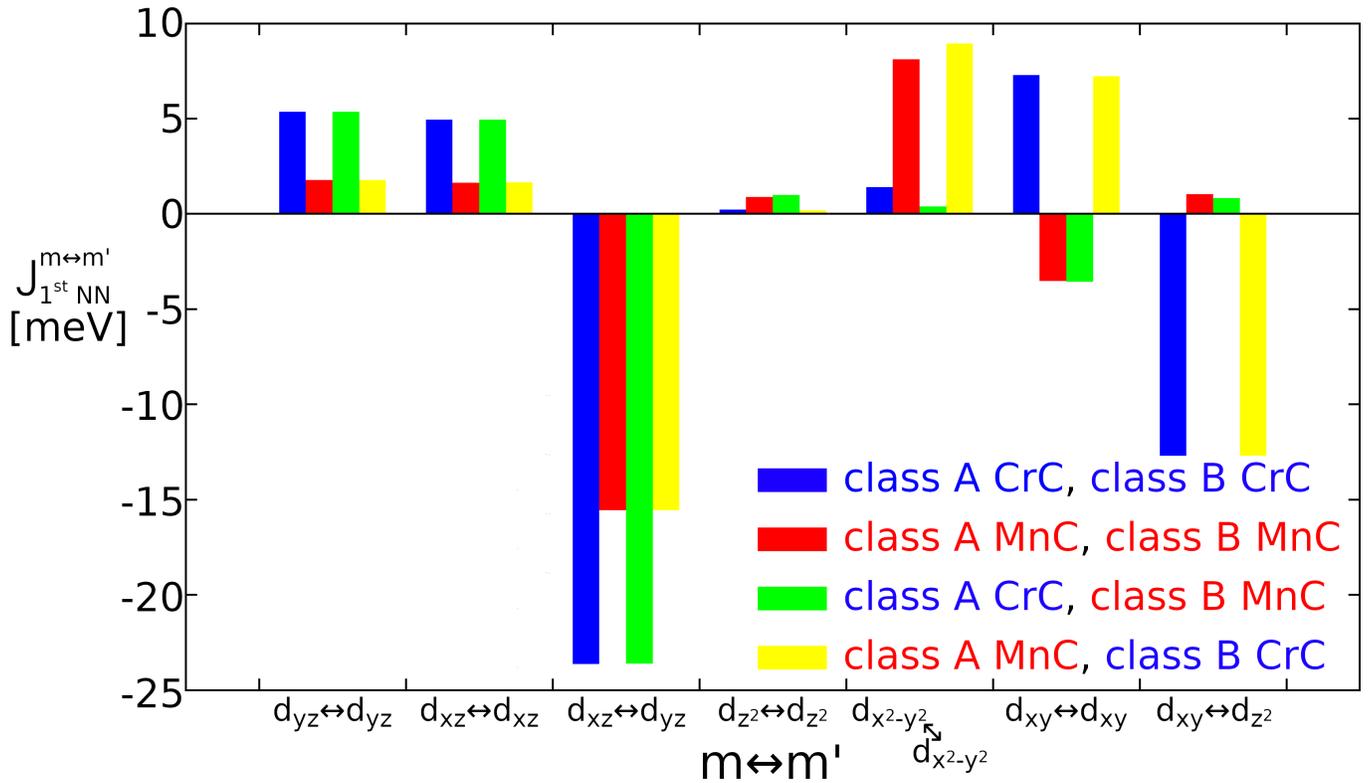

(a) Orbitally-resolved first-nearest-neighbor magnetic exchange.

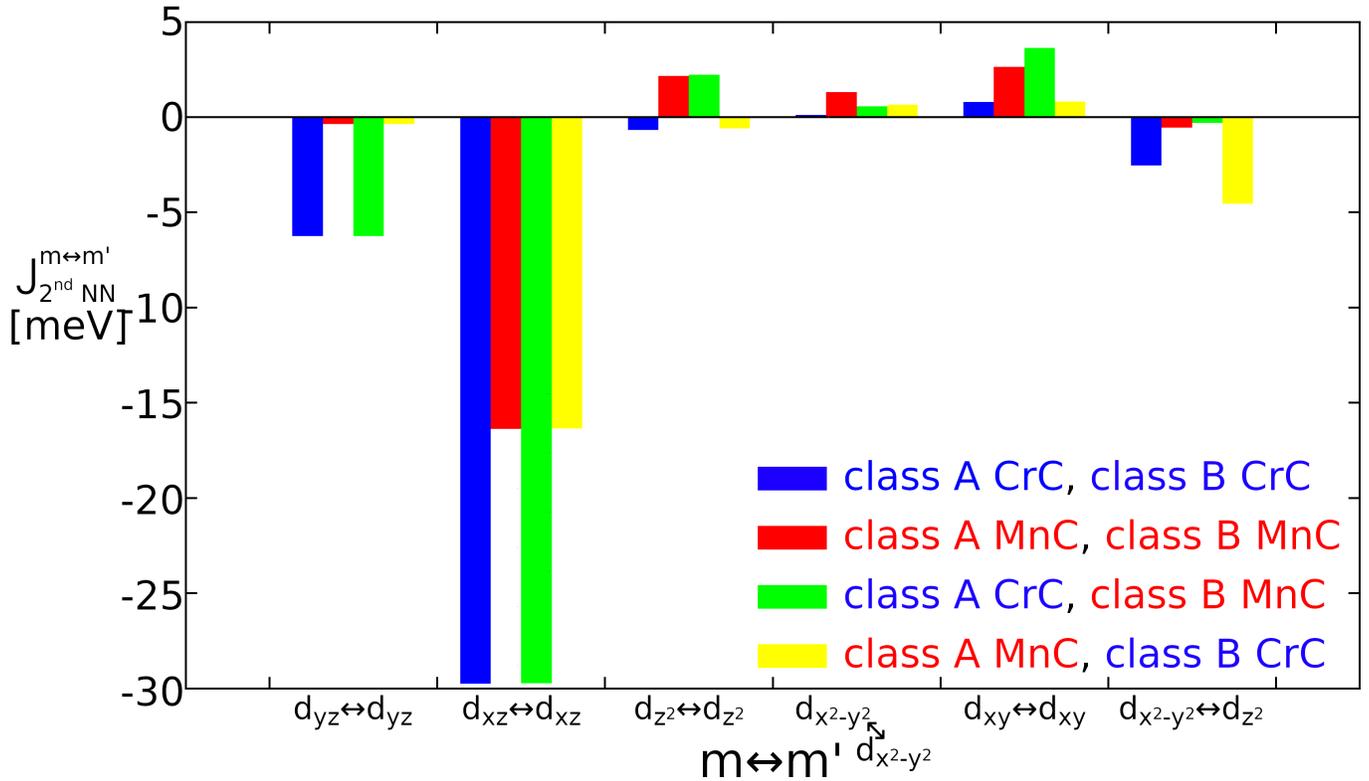

(b) Orbitally-resolved second-nearest-neighbor magnetic exchange.

FIG. 5: Combining the classes A and B from CrC and MnC results in the firm proof for class separation in both materials. In both cases of the first- and the second-nearest-neighbor "mixing" the results coincide with either pure CrC or pure MnC results, depending on where the corresponding classes are taken from.

*$d_\gamma - p_\sigma$ bond owing to a smaller overlap."* and that *"since the $d_\gamma - p_\sigma$ bond is stronger than the $d_\epsilon - p_\pi$ bond, the total superexchange interaction will nevertheless be antiferromagnetic."* In Fig. 5b we show that this is clearly not the case in CrC and MnC, and that dominant contribution $J_{d_{xz}-d_{xz}}$ involves one of the $p_\pi$ orbitals ($p_z$) and not the $p_\sigma$ ($p_x$).

To summarize, here we do not counter GKA rules on orbital level, on the contrary, our results agree with them. Nevertheless, we argue that the orbital-interaction mechanism which is considered non-important by GKA is crucial in two materials that we investigated, and vice versa - the mechanism considered dominant by GKA is actually negligible in our case.

### D. Direct- vs. super-exchange and the role of atomic environment

Finally, we continue with the competition between direct- and super-exchange, and the role of atomic environment on total magnetic exchange between two $d$ orbitals of neighboring M atoms. In order to discuss the influence of direct exchange, superexchange, and atomic environment to the orbitally-resolved components of the total magnetic interactions between the first- and the second-nearest-neighbor M atoms, we introduce three variables: variable D - representing direct hopping matrix elements between two $d$ orbitals on adjacent M atoms in DFT Hamiltonian; variable S - representing hopping matrix elements between $d$ orbitals from M, and $p$ orbitals from NM atom, i.e. the hopping connected with the superexchange; and variable E - representing the atomic environment and orbital occupation of M and NM atoms in both materials. For the sake of better understanding, we graphically present the atomic environments in MnC and CrC, in Figs. 6(a) and (b), respectively.

We start with the dominant contributions to the total first- and second-nearest-neighbor magnetic exchange, which originate from the class A orbitals, and in any case these contributions are strong FM. The similarity between the results for MnC and CrC is not surprising, since the three ingredients forming the magnetic exchange - D, S, and E - are almost the same in two materials.

Regarding the first-nearest-neighbor dominant contribution, one can note that direct exchange hopping elements between $d_{xz}$ on one magnetic atom and $d_{yz}$ on its first-nearest-neighbor (up and down) are 0.140 and 0.161 eV in CrC, and 0.125 and 0.124 eV in MnC. The magnitudes of hopping matrix elements connected with the superexchange process through the non-magnetic ligand ($d$-$p$ hopping matrix elements, up and down) are 0.999 and 1.382 eV in CrC and 0.714 and 1.230 eV in MnC. On top of everything the atomic environment is almost the same in two materials. By stating this, we refer to the fact that $d_{xz}$ and $d_{yz}$ orbitals on magnetic atoms, with spin up are occupied (energy of the orbital $< 0$) in both materials, while the same orbitals with the spin down are unoccupied (energy of the orbital $> 0$). In case of $p_z$ orbitals on C atom, the spin up orbitals are unoccupied in both materials, while $p_z$ orbitals with spin down are the only orbitals "seeing" different environment in two materials - in CrC this orbital is unoccupied, while in MnC it is just below (46 meV) the Fermi level. Situation is exactly the same in case of the second-nearest-neighbor magnetic exchange - all relevant factors are the same in two materials, including direct hopping between $d_{xz}$ and $d_{xz}$ orbitals on two second-nearest-neighbor M atoms, when their bond is parallel to the $x$ direction. In case of up spin, these matrix elements are 0.0018 and 0.0371 eV in CrC and MnC, respectively. For spin down, their magnitudes are 0.104 and 0.103 eV. In case of bond along $y$ direction, the important direct hopping is $d_{yz}$-$d_{yz}$, and its values are the same as those above, due to the 4-fold out-of-plane rotation symmetry of the system. To summarize, our calculations show that on the level of DFT, orbitals belonging to the class A behave almost the same in two materials and therefore produce almost the same orbital contributions to the magnetic exchange in them.

Now that the reasons for the similar dominant parts of the total exchange are elaborated, we continue with the secondary, yet more interesting contributions, originating from the states of class B. Those contributions are $J_{d_{xy}-d_{xy}}$ and $J_{d_{xy}-d_{z^2}}$ for the first-nearest-neighbor. Similarly as it was the case with the dominant contributions, here we identify three variables - D, S, and E. Regarding the D and S degrees of freedom of class B, the size of hopping elements with spin down are almost the same. The hopping elements connected with S are 1.215 and 1.294 eV, in CrC and MnC respectively, while those connected with D are 0.529 and 0.480 eV in CrC and MnC respectively. However, the magnitude of spin up hopping in CrC (S = 1.148 eV and D = 0.444 eV) is about 1.36/1.76 times the value from MnC (S = 0.843 eV and D = 0.252 eV). The similarities and differences between D and S hopping are also illustrated graphically in Figs. 2(i) and 3(i). Nevertheless, the most important difference seems to exist between the atomic environments in CrC and MnC. As one can see from Figs. 6 (a) and (b), the atomic environment for the class B states is drastically different in two materials - $d_{xy}$ orbital with spin up is 0.037 eV above Fermi level in case of CrC, while the same orbital is 3.897 eV below Fermi level in case of MnC. Another difference in atomic environment comes from $p_x$ and $p_y$ orbitals, regardless of the spin - in CrC they are all unoccupied, while in MnC they are all occupied. Regarding the second-nearest-neighbor magnetic exchange of the class B, the contributions are $J_{d_{xy}-d_{xy}}$ and $J_{d_{z^2}-d_{z^2}}$. Here the hopping elements connected with the superexchange are the same as in the case of the first-nearest-neighbors,

9

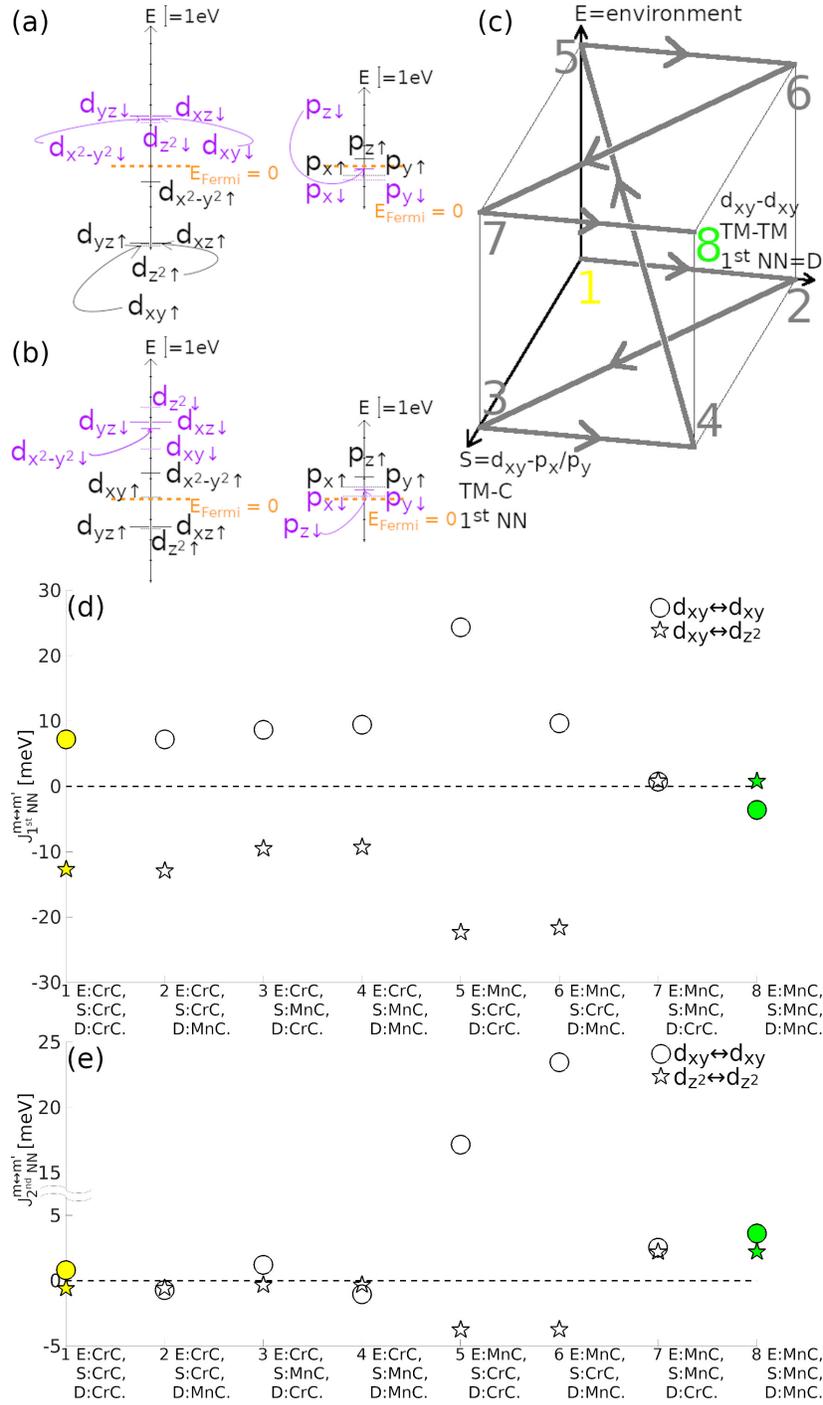

FIG. 6: The atomic environment and orbital occupation of (a) the Cr and C atoms in CrC, and (b) the Mn and C atoms in MnC. (c) The illustration of the "path" taken through the D-S-E space in order to examine the sensitivity of the secondary (class B) orbitally-resolved contribution w.r.t. all three degrees of freedom - D, S, and E. Yellow number 1 refers to the whole class B being from CrC, and green number 8 refers to the whole class B being from MnC. (d) The first-nearest-neighbor class B contributions w.r.t. D-S-E. (e) The second-nearest-neighbor class B contributions w.r.t. D-S-E.

while hopping elements connected with the direct exchange are all negligible (the largest one being the $d_{xy} - d_{xy}$ with spin up in MnC, and having value of 0.0436 eV). Therefore, the three important factors in case of both the first- and the second-nearest-neighbors are: D - the hopping between $d$ orbitals of the first-nearest-neighbor M atoms; S - the hopping between the $d$ orbital of M atom and $p$ orbital of its first-nearest-neighbor C atom; and E - the atomic environment which reflects the on-site energies and occupation of orbitals in class B. The visual representation of three-dimensional "space" formed by D, S, and E is given in Fig. 6(c). The gray line symbolizes the path that we took through this "space" - staring from all three (D, S, and E) being taken from CrC (yellow number 1) and finishing with all three being taken from MnC (green number 8). In Fig. 6(d) and (e) we present the results of the walk through the D-S-E space in case of the first- and the second-nearest-neighbor magnetic interactions between orbitals of the class B.

From Fig. 6(d) one can see that atomic environment has the largest effect on the orbital contributions to the total first-nearest-neighbor magnetic exchange. Namely, in case of atomic environment from CrC ("1"-"4" on the horizontal axis), one sees that $J_{d_{xy}-d_{xy}}$ is always positive (AFM), while $J_{d_{xy}-d_{z^2}}$ contribution is always negative (FM). Furthermore by comparing the "1"-"4", we see that next to the environment, the superexchange mechanism is important for both contributions, while direct exchange is the least important. When superexchange is from CrC, the value of $J_{d_{xy}-d_{xy}}$ is fixed to 7.2 meV, regardless of the direct exchange, and it jumps to around 9 meV if superexchange is taken from MnC, with the effect of direct exchange being 0.5 meV large in that case. The $J_{d_{xy}-d_{z^2}}$ contribution is fixed to $-12.8$ meV when superexchange is taken from CrC, and to $-9.3$ meV in case that superexchange is from MnC. As one could expect, the direct $d_{xy} - d_{xy}$ interaction has no effect on $J_{d_{xy}-d_{z^2}}$ contribution. However, in case of environment from MnC ("5"-"8" on the horizontal axis) the situation is not as straightforward as it was in case of environment from CrC. On one hand, for the contribution $J_{d_{xy}-d_{z^2}}$ situation is the same as in CrC environment. This contribution is mainly determined by superexchange, and it is exactly the superexchange from MnC (see star symbols at points "7" and "8") that makes this contribution weakly AFM (0.8 meV). If it was for superexchange from CrC, this contribution would be strongly FM ($-21$ meV, see star symbols at points "5" and "6"), rather than weak AFM. On the other hand, the behavior of $J_{d_{xy}-d_{xy}}$ is drastically different. Namely, in case of the MnC environment, both direct- and super-exchange seem to play the important role, with superexchange still dominating - in case of superexchange being from CrC, the contribution is strongly AFM ($10-25$ meV, see circles at points "5" and "6"), whereas in case of superexchange from MnC the contribution is weak and changes sign with direct exchange (see circles at points "7" and "8"). Therefore, one could conclude that $J_{d_{xy}-d_{xy}}$ contribution in case of the first-nearest-neighbor tends to always be AFM, and the MnC is exception, due to the coincidence of particular D, S, and E.

In case of the second-nearest-neighbor magnetic exchange, $J_{d_{z^2}-d_{z^2}}$ is again mainly determined by the environment, and secondly by the superexchange. In case of the environment from MnC, the effect of superexchange is more prominent - it changes the sign of orbital contribution when taken from two materials (see star symbols at "5", "6" and "7", "8" coordinates of Fig. 6(e)) - while in case of the environment from CrC, it only shifts the contribution from $-0.6$ ("1" and "2") to $-0.3$ meV ("3" and "4"). As one could anticipate, the direct exchange $d_{xy} - d_{xy}$ between the first-nearest-neighbor M atoms has no effect on $J_{d_{z^2}-d_{z^2}}$ between the second-nearest-neighbor M atoms. However, the other contribution to the second-nearest-neighbor magnetic exchange, $J_{d_{xy}-d_{xy}}$, hides a surprise. Namely, in case of the atomic environment of the CrC, this contribution is mainly determined by the direct exchange - $d_{xy} - d_{xy}$ - between the first-nearest-neighbor M atoms. In case D is taken from CrC, $J_{d_{xy}-d_{xy}}$ is AFM, and in case D is taken from MnC, $J_{d_{xy}-d_{xy}}$ is FM. Therefore, we conclude that the leading mechanism in case of the CrC environment is the unusual superexchange which is, unlike the usual one, mediated by the M atom instead the NM atom - the mechanism we did not even consider to exist based on available literature. In case of the MnC environment, the usual superexchange dominates over the unusual one. It determines $J_{d_{xy}-d_{xy}}$ to be strongly AFM ($17-23$ meV) when superexchange is taken from CrC, and weakly AFM ($2.5-3.6$ meV) when the superexchange is taken from MnC.

To summarize this section, the first-nearest-neighbor magnetic exchange in CrC is consequence of certain atomic environment and the superexchange, while in MnC the $J_{d_{xy}-d_{xy}}$ is due to the coincidence of all three important factors - D, S and E - and if one is changed, the interaction changes sign. Regarding the second-nearest-neighbor magnetic exchange, in CrC the $J_{d_{xy}-d_{xy}}$ is determined by the unusual superexchange mediated by the magnetic atom, while in MnC, the usual superexchange is responsible for the sign and magnitude of the contributions of class B states.

### E. Third-nearest-neighbor magnetic exchange - $d_{xy}$-$d_{xy}$ contribution

In this section we briefly address magnetic exchange between $d_{xy}$-$d_{xy}$ orbitals on two third-nearest-neighbor M atoms. As it was the case with the second-nearest-neighbor exchange, here we have negligible direct hopping between two $d_{xy}$ orbitals (the largest magnitude of such hopping is 0.02 eV for spin-down interaction in MnC). Therefore, the

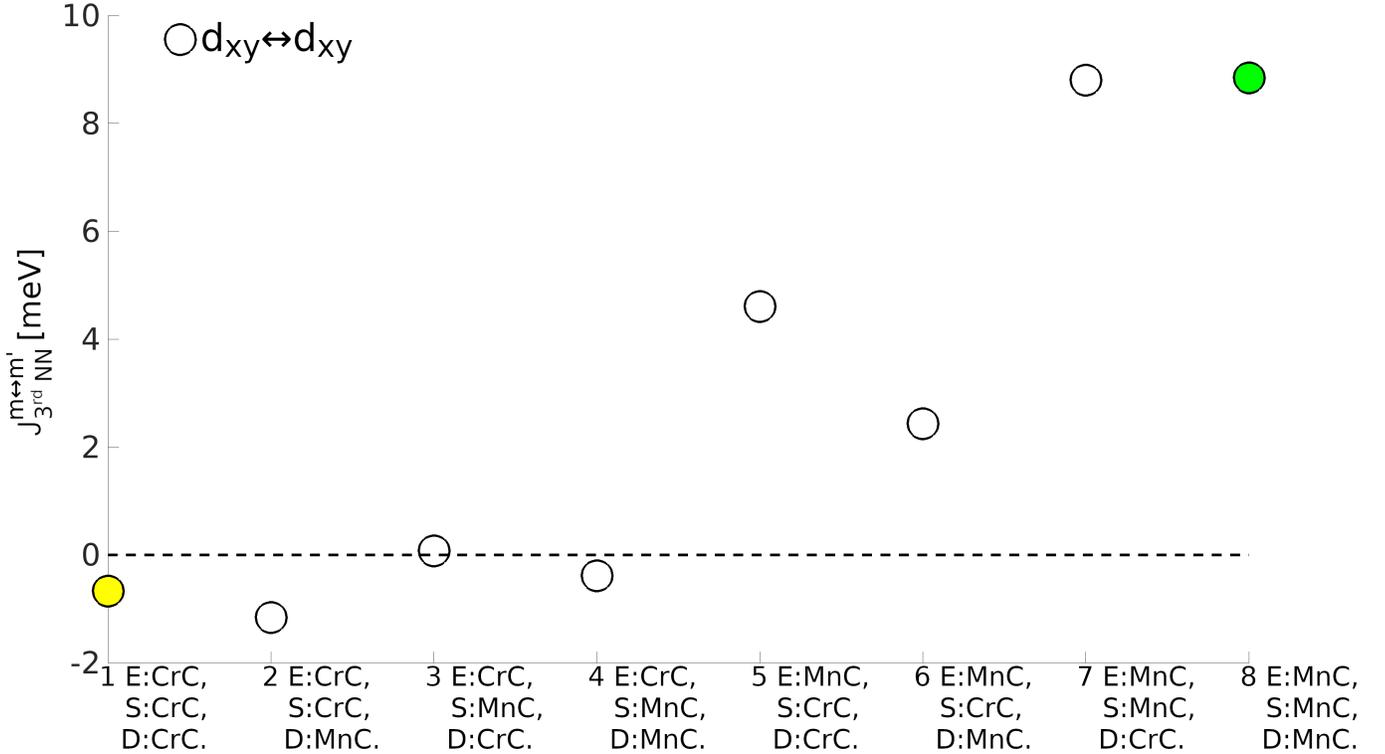

FIG. 7: The third-nearest-neighbor class B contributions w.r.t. D-S-E.

third-nearest-neighbor interaction can originate only from usual (S) or unusual (D) superexchange. From the Fig. 7 one can draw two conclusions: (1) that in case of the environment of CrC it is unclear which mechanism is dominant, since changing any of the two introduces significant relative change, and (2) that in case of the environment of MnC the usual superexchange (S) through the non-magnetic ligands determines the strength of the magnetic exchange between $d_{xy}$-$d_{xy}$ orbitals on two third-nearest-neighbor M atoms.

A careful reader would note that in both materials the sign of the first- and the third-nearest-neighbor differ. However, due to the fact that different mechanisms dominate in two materials, in case of the first- and the third-nearest-neighbor, this is not surprising. In case of the first-nearest-neighbor exchange between two $d_{xy}$ orbitals, in CrC the leading mechanism is superexchange (S) through the NM ligand, while in MnC it is unclear whether it is direct- or super-exchange. In case of the third-nearest neighbor it is exactly opposite - in CrC the leading mechanism is unknown, and in MnC it is usual superexchange (S) through the NM ligand. Even though it is not surprising that signs may differ for different nearest-neighbor interactions in any of the two materials, this is not a general rule. If one compares the open dot data in Fig. 6(d) (the first-nearest-neighbor) and in the Fig. 7 (the third-nearest-neighbor), those have the same sign in four cases ("3", "5", "6", "7") and different in four other cases ("1", "2", "4", "8"). This "50-50" division confirms that difference in sign between the first- and the third-nearest-neighbor is not a general rule for the class of CrC-MnC-like materials, but merely a coincidence.

## APPENDIX A: SYMMETRY CONSTRAINTS

Within this Appendix we derive the symmetry imposed constraints on the magnetic exchange tensor for the first nearest neighbor, in case of a general flat material with square lattice and inversion center between TM ions. The inversion center forces magnetic exchange to be symmetric, i.e. $\mathbf{J}_{A,B}^{pq} = \mathbf{J}_{A,B}^{qp}$, where $p, q \in x, y, z$ [19]. Further, the underlying idea is to use the fact that system possess the xy mirror plane, other two mirror planes ($M_2$ and $M_3$ on Fig. 8) and 4-fold rotational symmetry around z axis, and derive corresponding symmetry rules. The xy mirror plane symmetry results in:

$$\mathbf{J}_{O,A} = \mathbf{M}_{xy} \cdot \mathbf{J}_{O,A} \mathbf{M}_{xy}{}^T. \tag{1}$$

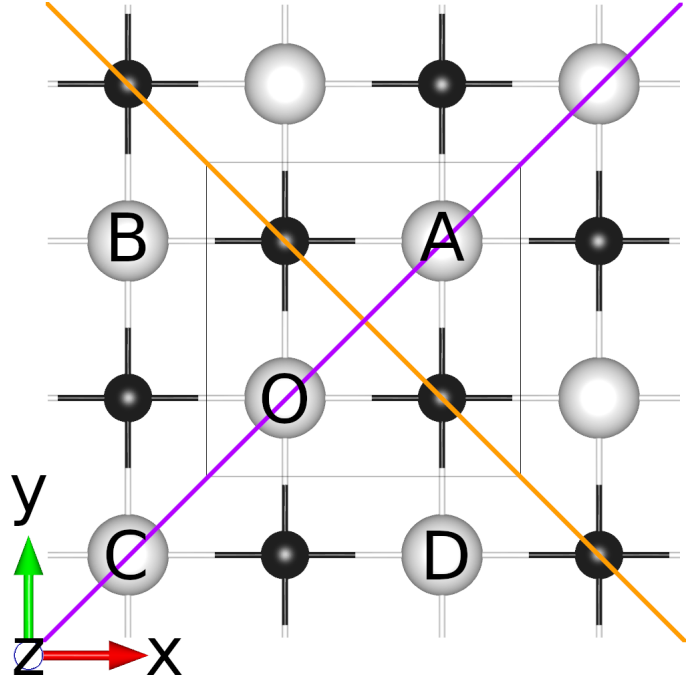

FIG. 8: Top view of flat, square lattice TMX system, with the first nearest TM neighbors labeled with O, A, B, C, and D. Mirror planes of the system are indicated with purple ($M_2$) and orange ($M_3$).

where $\mathbf{M}_{xy}$ is the matrix representation of the mirror plane effect on Cartesian basis vectors. Its matrix form is:

$$\mathbf{M}_{xy} = \begin{bmatrix} 1 & 0 & 0 \\ 0 & 1 & 0 \\ 0 & 0 & -1 \end{bmatrix}. \tag{2}$$

After performing matrix multiplication on the right hand side and equalizing it with the left hand side, one obtains $J^{xz} = -J^{xz} = 0$ and $J^{yz} = -J^{yz} = 0$. Other matrix elements are not restricted by this symmetry. Using the symmetry of any of the two, $M_2$ or $M_3$, will result in single additional limitation of $J^{xx} = J^{yy} = J$. Finally, the use of the 4-fold rotational symmetry around z axis results in the fact that matrices $\mathbf{J}_{O,A}$ and $\mathbf{J}_{O,B}$ have opposite xy elements $J^{xy}_{O,A} = -J^{xy}_{O,B}$. This brings us to the final conclusion that in the most general form, the exchange tensors of flat, square lattice system can all be written down in terms of three parameters ($J$, $J^{xy}$, and $J^{zz}$) as:

$$\mathbf{J}_{O,A} = \mathbf{J}_{O,C} = \begin{bmatrix} J & J^{xy} & 0 \\ J^{xy} & J & 0 \\ 0 & 0 & J^{zz} \end{bmatrix}, \tag{3}$$

and

$$\mathbf{J}_{O,B} = \mathbf{J}_{O,D} = \begin{bmatrix} J & -J^{xy} & 0 \\ -J^{xy} & J & 0 \\ 0 & 0 & J^{zz} \end{bmatrix}. \tag{4}$$

For purpose of our material screening, we have considered mean-field-like case where $J^{xy}$ and $-J^{xy}$ from two different the first-nearest-neighbors cancel out and nearest neighbor tensors are purely diagonal. Furthermore, this assumption is justified - and later in our study confirmed in thorough analysis using TB2J - due to the small spin-orbit coupling in C and consequent negligible anisotropy in exchange.

## APPENDIX B: SCREENING PARAMETER EXTRACTION

Here we derive the equations for obtaining magnetic parameters from DFT energies. Starting from Heisenberg Hamiltonian for this system:

$$H = \sum_{i,j} \mathbf{S}_i \begin{bmatrix} J & 0 & 0 \\ 0 & J & 0 \\ 0 & 0 & J+\delta \end{bmatrix} \mathbf{S}_j + \sum_i \mathbf{S}_i \begin{bmatrix} 0 & 0 & 0 \\ 0 & 0 & 0 \\ 0 & 0 & A \end{bmatrix} \mathbf{S}_i, \tag{5}$$

we write the energies (per primitive unit cell) of the four characteristic magnetic configurations: FM in-plane, FM out-of-plane, AFM in-plane, and AFM out-of-plane.

$$E_{FMin} = 4 \cdot J S^2, \tag{6}$$

$$E_{FMout} = 4 \cdot (J+\delta) S^2 + 2 \cdot A S^2, \tag{7}$$

$$E_{AFMin} = -4 \cdot J S^2, \tag{8}$$

$$E_{AFMout} = -4 \cdot (J+\delta) S^2 + 2 \cdot A S^2. \tag{9}$$

By combining the equations (6) and (8), one can easily extract

$$J_{in-plane} = J S^2 = \frac{E_{FMin} - E_{AFMin}}{8}. \tag{10}$$

Further, combining (6) and (7) into one equation, and (8) and (9) into another, results in:

$$E_{FMout} - E_{FMin} = 4 \cdot \delta S^2 + 2 \cdot A S^2, \tag{11}$$

and

$$E_{AFMout} - E_{AFMin} = -4 \cdot \delta S^2 + 2 \cdot A S^2. \tag{12}$$

Now, it is straight-forward to extract the desired parameters, $\Delta = \delta S^2$ and $A^{zz} = A S^2$:

$$\begin{aligned} \Delta &= \frac{(E_{FMout} - E_{FMin}) - (E_{AFMout} - E_{AFMin})}{8} \\ &= \frac{E_{FMout} - E_{AFMout}}{8} - \frac{E_{FMin} - E_{AFMin}}{8} \\ &= J_{out-of-plane} - J_{in-plane}, \end{aligned} \tag{13}$$

and

$$A^{zz} = \frac{(E_{FMout} - E_{FMin}) + (E_{AFMout} - E_{AFMin})}{4}. \tag{14}$$

Note that the magnitude of the local magnetic moments is incorporated in the interaction parameters, and hence vectors $\vec{S}_i$ are unit vectors along the direction of local magnetic moments at the site $i$.


\fi
\end{document}